\documentclass[usenatbib]{mn2e}
\usepackage{gensymb}
\usepackage{graphicx,amsmath}
\def \s{\hphantom{0}}
\def\apj{\rm ApJ}
\def\apjl{\rm ApJL}
\def\apjs{\rm ApJS}
\def\aj{\rm AJ}
\def\mnras{\rm MNRAS}
\def\nat{\rm Nature}

\def\pasp{\rm PASP}
\def\aap{\rm AAP}
\def\araa{\rm ARA\&A}

\def\prd{\rm PRD}
\def\apss{\rm Ap\&SS}

\def\gax{\mathrel{\raise.3ex\hbox{$>$}\mkern-14mu\lower0.6ex\hbox{$\sim$}}}
\def\lax{\mathrel{\raise.3ex\hbox{$<$}\mkern-14mu\lower0.6ex\hbox{$\sim$}}}
\def\gtorder{\mathrel{\raise.3ex\hbox{$>$}\mkern-14mu
             \lower0.6ex\hbox{$\sim$}}}
\def\ltorder{\mathrel{\raise.3ex\hbox{$<$}\mkern-14mu
             \lower0.6ex\hbox{$\sim$}}}

\begin{document}

\title[First Candidates for Failed Supernovae]
{The Search for Failed Supernovae with The Large Binocular Telescope: First Candidates}

\author[J. R. Gerke et al.]{J.~R.~Gerke$^1$, C. S. Kochanek$^{1,2}$, K. Z. Stanek$^{1,2}$ \\
  $^{1}$Department of Astronomy, The Ohio State University, 140 West 18th Avenue, Columbus OH 43210 \\
  $^{2}$Center for Cosmology and AstroParticle Physics, The Ohio State University, 191 W. Woodruff Avenue, Columbus OH 43210 \\
}

\maketitle

\begin{abstract}
We are monitoring 27 galaxies within 10 Mpc 
using the Large Binocular Telescope 
to search for failed supernovae (SNe), massive stars that collapse to form a black 
hole without a SN explosion. 
We present the results from the first 4 years of survey data, during which these 
galaxies were observed to produce 3 successful core-collapse SNe. 
We search for 
stars that have ``vanished'' over the course of our survey, by examining all stars 
showing a decrease in luminosity of $\Delta \nu L_{\nu} \ge 10^4L_{\odot}$ from the 
first to the last observation.  
We also search for the low luminosity, long duration transients 
predicted by \citet{lovegrove2013} for failed explosions of red supergiants. 
After analyzing the first 4 years of data in this first direct search for failed SNe, 
we are left with one candidate requiring further study.  
This candidate has an estimated mass of 18-25$M_{\odot}$, a mass range likely 
associated with failed SNe and, if real, implies 
that failed SN represent a median fraction
of $f \simeq 0.30$ of core-collapses, with symmetric 90\% confidence limits of 
$0.07 \le f \le 0.62$. If follow up data eliminate this candidate, we find 
an upper limit on the fraction of core collapses leading to a failed SN of 
$f<0.40$ at 90\% confidence. 
As the duration of the survey continues to increase, it will begin to constrain the 
$f \simeq$ 10-30\% failure rates needed to explain the deficit of massive SN progenitors 
and the observed black hole mass function. 
\end{abstract}


\section{Introduction}
Massive stars have a tremendous impact on the evolution of galaxies and stellar
populations through both energy injection and chemical 
enrichment.  To understand how massive stars affect their environments, we 
must understand their deaths, particularly the 
balance between successful supernovae (SNe), which eject most of their mass, including 
significant amounts of nuclear processed material to be recycled into a new generation 
of stars, and failed SNe, which form a black hole while ejecting little or no energy or 
enriched material (see \citealt{smartt2009a} for a review of SN progenitors).  
Understanding this balance is also crucial to 
understanding the SN 
mechanism, a long standing problem in astronomy (see \citealt{ondrej2014}).  

It has long been believed that some fraction of massive stars experience a failed SN.  The focus has 
usually 
been on high mass stars at lower metallicity where mass loss may be smaller \citep{heger2003}.  
However, there are also arguments about whether elemental abundances require a significant 
fraction of failed SNe for M $\ga$ 25 $M_{\odot}$ to avoid the overproduction of heavy 
elements (e.g. \citealt{maeder1992} for and \citealt{prantzos1994} against).  
\citet{brown2013} find abundances are fit well with both no failed SNe 
and having all stars above 25 $M_{\odot}$ experience a failed SN.  Pushing the failed SN limit
down to 18 $M_{\odot}$ requires a star formation rate 3 times the fiducial value.  However, 
they do not explore the consequences of failed SNe from a more complex range of masses other than 
a simple mass limit.  \citet{clausen2014} used a more physical model for the mass distribution 
of failed SNe and found only weak constraints on any mass range 
associated with failed SNe from abundance measurements. 

The question of failed SNe has become more pressing because there appears to be 
a dearth of high mass SN progenitors \citep{kochanek2008}. In particular, 
\citet{smartt2009} find no Type IIP progenitors
 more massive than those corresponding to initial masses of $\simeq$ 18 $M_{\odot}$, 
even though stars with initial masses up to 
 25-30 $M_{\odot}$ are thought to explode as red supergiants (RSG).  
It is possible that this RSG problem has a solution other than failed SNe, 
such as different physical treatments in the stellar evolution models, as discussed 
by \citet{smartt2009}. For example, the rotating models of \citet{groh2013} 
can move the upper mass limit to explode as a red supergiant to as low as 
17 $M_{\odot}$, although these models may be in conflict with recent astroseismic 
results \citep{ceillier2012}.
Another option is to make these stars very dusty \citep{walmswell2012}, so that they 
would be fainter and thus have underestimated masses if the extinction 
is not well understood, although, these particular 
models overestimate the net effect (see \citealt{kochanek2012a}).  
While the progenitor masses depend somewhat on the analysis (e.g., \citealt{maund2014}), the basic 
conclusion of \citet{kochanek2008} and \citet{smartt2009} appears to be robust.  
Moreover, studies using local stellar populations to estimate progenitor masses 
of historical SNe have found similar results (\citealt{jennings2012}; \citealt{williams2014}; \citealt{jennings2014}).
Failed SNe in this mass range can also explain the black hole mass function \citep{chris2014,kochanek2014b} 
and the mismatch between star formation and SN rate estimates (\citealt{horiuchi2011}, 2014).

We estimate that the failed ccSNe fraction $f$ associated with
the red supergiant problem is $f=0.20$ with a crude 90\%
confidence range of $0.11 < f < 0.33$.  We made this estimate
by modeling the upper and lower mass limits for RSG core
collapses using a Gaussian model for the Smartt et al. (2009)
results, a uniform distribution from $25$ to $30 M_\odot$ for
the upper mass limit of stars that are RSGs at death, and a
uniform distribution from $100$ to $200 M_\odot$ for the
upper mass limit of the IMF.   If we disallow core collapse 
below $8$-$9M_\odot$, the lower limit
rises to $f \simeq 0.15$.   This range is very similar to the
estimates of $f=0.18$ ($0.09 < f < 0.39$) associated with
explaining the black hole mass function using failed ccSNe
(Kochanek 2015), although these limits fixed the minimum
mass for core collapse to $9M_\odot$ and allow failed
ccSNe of stripped stars as well as RSGs.

While simulations of SNe cannot at present predict which SN will succeed, they can explore 
which stars are more difficult to explode given their mass and internal structure. 
\citet{oconnor2011} used the compactness of the core at bounce to estimate 
whether a SN is likely to be successful.  They find that progenitors in the mass range associated 
with the RSG problem also have structures that make them more difficult to explode.
\citet{ugliano2012} also found that this progenitor mass range is more likely 
to result in a failed SN. In their study of the neutrino mechanism for SNe, 
\citet{ondrej2014} show that a number of mass ranges probably lead 
to failed SNe, again including the mass range associated with the RSG problem, 
and that a failure rate $f \simeq 20-30\%$ is quite probable.  This is also 
supported by the black hole mass function \citep{chris2014,kochanek2014b}.

Despite the expectation that some massive stars end their lives in a failed SN, 
there are surprisingly few studies of the external appearance of such events.  
 \citet{woosley2012} found that some stars can collapse without fallback, probably 
forming a black hole without a significant transient.
Failed SNe of RSGs probably produce a low luminosity optical transient.  
\citet{lovegrove2013}, motivated by \citet{nadezhin1980}, 
simulated failed SNe of $15$ and $25M_{\odot}$ RSGs, finding that the mass energy lost 
in neutrinos leads to a weak shock that unbinds the stellar envelope.  
The resulting black hole has the mass of the helium core, and this would naturally explain 
the compact remnant mass function (\citealt{chris2014,kochanek2014b}; \citealt{clausen2014}).
The optical signature of such failed SNe is a shock breakout followed by a 
longer term transient.  The shock breakout 
produces a brighter 
(few $10^7L_{\odot}$) optical transient, but it only lasts for 3-10 days \citep{piro2013}. 
The longer term transient is driven by the recombination of the unbound envelope of the star, 
leading to a transient that lasts 
about $\sim1$~year with a luminosity of order $L_{bol} \sim 10^6L_{\odot}$ and a temperature 
of $\sim4000$~K at peak \citep{lovegrove2013}.  While conditions are ideal for dust 
formation, dust formation only occurs as the transient begins to fade and so affects the 
optical signature little \citep{kochanek2014c}.  
No matter what the intervening physics, the star must ultimately ``vanish'' 
in the optical (see \citealt{kochanek2008}). 

Traditionally, the search for failed SNe has focused on neutrinos 
(e.g., \citealt{lien2010}; \citealt{muhlbeier2013}) or gravitational waves 
(e.g., \citealt{ott2009}; \citealt{kotake2011}) with the difficulty that 
they could only be detected in the Milky Way with event 
timescales of centuries for a Galactic SN rate of 1 per 50 to 100 years (see \citealt{adams2013}).  
However, in \citet{kochanek2008} we pointed out that an optical survey for failed SNe 
was possible and could explore many more galaxies than simply the Milky Way.  
With an observed SN rate in the sample of $\sim1$ per year, the time scales to detect a 
failed SN become much more practical.  In essence, 
we are monitoring the health of $\sim10^6$ RSGs in 27 galaxies 
within 10 Mpc using the Large Binocular Telescope (LBT) to see if any die, independent 
of the external symptoms beyond ultimately vanishing.  
With such a large sample of massive stars, we are able to probe the expected rates of this 
rare, difficult to observe phenomenon much more rapidly than if examining only 
our Galaxy.  

In this work we report on the first such observational search for failed SNe.  
In \S2 we discuss the galaxy sample and survey observations 
while \S3 explains our image subtraction and calibration methods.  In \S4
we outline the candidate selection process and in \S5 discuss the successful SNe 
in our sample.  We discuss the last three rejected candidates in \S6.  In \S7 we detail our 
remaining candidate and in \S8 we place limits on the failed SN rate.  
We discuss future directions in \S9 and summarize our results in \S10. 

\section{Galaxy Sample and Observations}
Our galaxy sample consists of 27 galaxies within 10 Mpc, essentially those selected 
in \citet{kochanek2008} that are visible from the 
LBT \citep{hill2006} and their nearby companions.  We do not include 
M31 and M33 because their sizes are poorly matched to our instrument.  
The galaxies and several of their 
properties are given in Table~\ref{tab:galsam}. 
The observing strategy made use of the LBT's unique binocular feature, observing 
in the $R$ Bessel filter with the red-optimized LBC-Red camera while simultaneously cycling 
through observations in the $U_{spec}$ interference filter and the $B$ and $V$ Bessel filters with the blue optimized LBC-Blue 
camera \citep{giallongo2008}.  We calibrate our photometry to the Johnson-Cousins system, therefore 
we will refer to our calibrated R band magnitudes as $R_c$ to differentiate from the filter. 
Additionally, the $U_{spec}$ interference filter is very similar to the Bessel filter and we 
will refer to the calibrated magnitudes as $U$ band. 
Each LBC camera consists of four 4096 $\times$ 2048 CCDs, each of which covers 
$17.3 \times 7.7$ arcmi$\mbox{n}^2$ with a plate scale of $0\farcs225$ per pixel.  Chips
1, 2 and 3 are adjacent to each other along the long sides of the chips, 
while Chip 4 is above them and rotated by 90$^\circ$ (see Figure 4 in \citealt{giallongo2008}), 
to provide a total area of roughly $23\arcmin \times 23\arcmin$.  
The majority of the target galaxies fit on the central chip, Chip 2. 
The larger galaxies required multiple chips:  M81 (all chips), 
M101 (all chips), NGC~2403 (all chips), NGC~6946 (3 chips), NGC~628 (3 chips), 
NGC~4258 (2 chips) and NGC~672 (2 chips).  A single pointing included
NGC~3627 on Chip 3 and NGC~3628 on Chip 1. Another single pointing included
NGC~4258 on Chip 2 and NGC~4248 on Chip 4. 

\begin{table*}
\caption{Galaxy Sample}
\label{tab:galsam} 
\begin{tabular}{lcccccccc}
\hline \hline
  Galaxy &  Distance &  Number of &  \multicolumn{2}{c}{Observation} &  Baseline &    $\nu L_{\nu}/L_{\odot}$ per  & Depth & Distance \\
         & (Mpc)     & Epochs     & First & Last & (years) & Count & $R_c$ mag & Reference\\
\hline
M81 &  \s3.65 &   24 & 2008-03-08 & 2013-01-09     & 4.8 & 0.50   & 25.48 &  1 \\
M82 &  \s3.52 &    16 & 2008-03-08 & 2013-01-10    & 4.8 & 1.11   & 25.43 & 2 \\
M101 &  \s6.43 &   13 & 2008-03-08 & 2013-01-10    & 4.8 & 0.49   & 25.58 & 3 \\
NGC~628 &  \s8.59 &   11 & 2009-01-31 & 2013-01-09 & 3.9 & 1.36   & 25.70 & 4 \\
NGC~672 &  \s7.20 &   12 & 2008-07-05 & 2013-01-09 & 3.2 & 0.87   & 26.16 & 5 \\
NGC~925 &  \s9.16 &   12 & 2008-07-06 & 2013-01-09 & 3.2 & 1.44   & 26.16 & 6 \\
NGC~2403 & \s3.56 &  23 & 2008-05-05 & 2013-01-10  & 3.9 & 0.48   & 26.24 & 7 \\
NGC~2903 & \s8.90 & \s9 & 2008-03-08 & 2013-01-09  & 4.7 & 1.04   & 25.49 & 8 \\
NGC~3077 & \s3.82 &  10 & 2008-05-04 & 2013-01-10  & 3.9 & 0.48   & 26.10 & 5 \\
NGC~3344 & \s6.90 & \s7 & 2008-05-04 & 2012-03-23  & 3.9 & 0.84   & 26.23 & 9 \\
NGC~3489 & \s7.18 & \s6 & 2008-03-12 & 2012-03-23  & 4.0 & 1.14   & 26.28 & 10 \\
NGC~3627 & 10.62 &  \s7 & 2008-05-04 & 2012-04-28  & 3.2 & 3.32   & 25.43 & 11 \\
NGC~3628 & 10.62 &  \s7 & 2008-05-04 & 2012-04-28  & 3.2 & 3.48   & 25.43 & 11 \\
NGC~4214 & \s2.98 & \s5 & 2008-03-13 & 2013-01-10  & 3.8 & 0.29   & 25.60 & 12 \\
NGC~4236 & \s3.65 & \s7 & 2008-03-09 & 2013-01-10  & 3.9 & 0.34   & 26.10 & 1 \\
NGC~4248 & \s7.21 &  28 & 2008-03-08 & 2013-01-10  & 4.8 & 1.09   & 25.60 & 13 \\
NGC~4258 & \s7.21 &  28 & 2008-03-08 & 2013-01-10  & 4.8 & 1.09   & 25.60 & 13 \\
NGC~4395 & \s4.27 & \s4 & 2008-03-10 & 2013-01-09  & 1.7 & 0.44   & 25.60 & 14 \\
NGC~4449 & \s3.82 & \s6 & 2008-03-09 & 2013-01-09  & 4.8 & 0.32   & 25.70 & 15\\
NGC~4605 & \s5.47 & \s5 & 2008-03-13 & 2013-01-10  & 2.1 & 0.53   & 25.72 & 16 \\
NGC~4736 & \s5.08 & \s5 & 2008-03-10 & 2013-01-09  & 3.9 & 1.19   & 26.23 & 17 \\
NGC~4826 & \s4.40 & \s6 & 2008-03-08 & 2013-01-10  & 4.7 & 0.35   & 26.23 & 2 \\
NGC~5194 & \s8.30 & \s8 & 2008-03-09 & 2013-01-10  & 3.9 & 0.93   & 26.16 & 18 \\
NGC~5474 & \s6.43 & \s7 & 2008-03-13 & 2013-01-10  & 3.9 & 0.60   & 26.16 & 3 \\
NGC~6503 & \s5.27 & \s7 & 2008-05-04 & 2012-10-15  & 4.3 & 0.70   & 26.16 & 6 \\
NGC~6946 & \s5.96 &  19 & 2008-05-03 & 2012-10-17  & 4.5 & 0.16   & 25.94 & 19 \\
IC~2574 & \s4.02 & \s9 & 2008-03-13 & 2013-01-10   & 4.7 & 0.46   & 26.06 & 6 \\
\hline
\multicolumn{9}{p{0.8\textwidth}}{The baseline is the time from the second 
observation to the last observation in the selection period.  The flux 
calibration $\nu L_{\nu}/L_{\odot}$ gives a sense of the number of counts expected from 
a star of a given luminosity.  The depth is the LBC/ETC estimate of $SNR=5$ for the $R_c$ band 
depth of the observations in an extragalactic field (i.e. a normal sky background).  
References -- (1) \citealt{gerke2011}; 
(2) \citealt{jacobs2009}; (3) \citealt{shappee2011}; (4) \citealt{herrmann2008}; 
(5) \citealt{karachentsev2004}; (6) \citealt{karachentsev2003}; (7) \citealt{willick1997}; 
(8) \citealt{drozdovsky2000}; (9) \citealt{verdes2000}; (10) \citealt{theureau2007}; 
(11) \citealt{kanbur2003}; (12) \citealt{dopita2010}; 
(13) \citealt{herrn1999}; (14) \citealt{thim2004}; 
(15) \citealt{annibali2008}; (16) \citealt{karachentsev2006}; 
(17) \citealt{tonry2001}; (18) \citealt{poznanski2009}; 
(19) \citealt{karachentsev2000}.
}
\end{tabular}
\end{table*}

\begin{figure}
  \centerline{\includegraphics[width=3.5in]{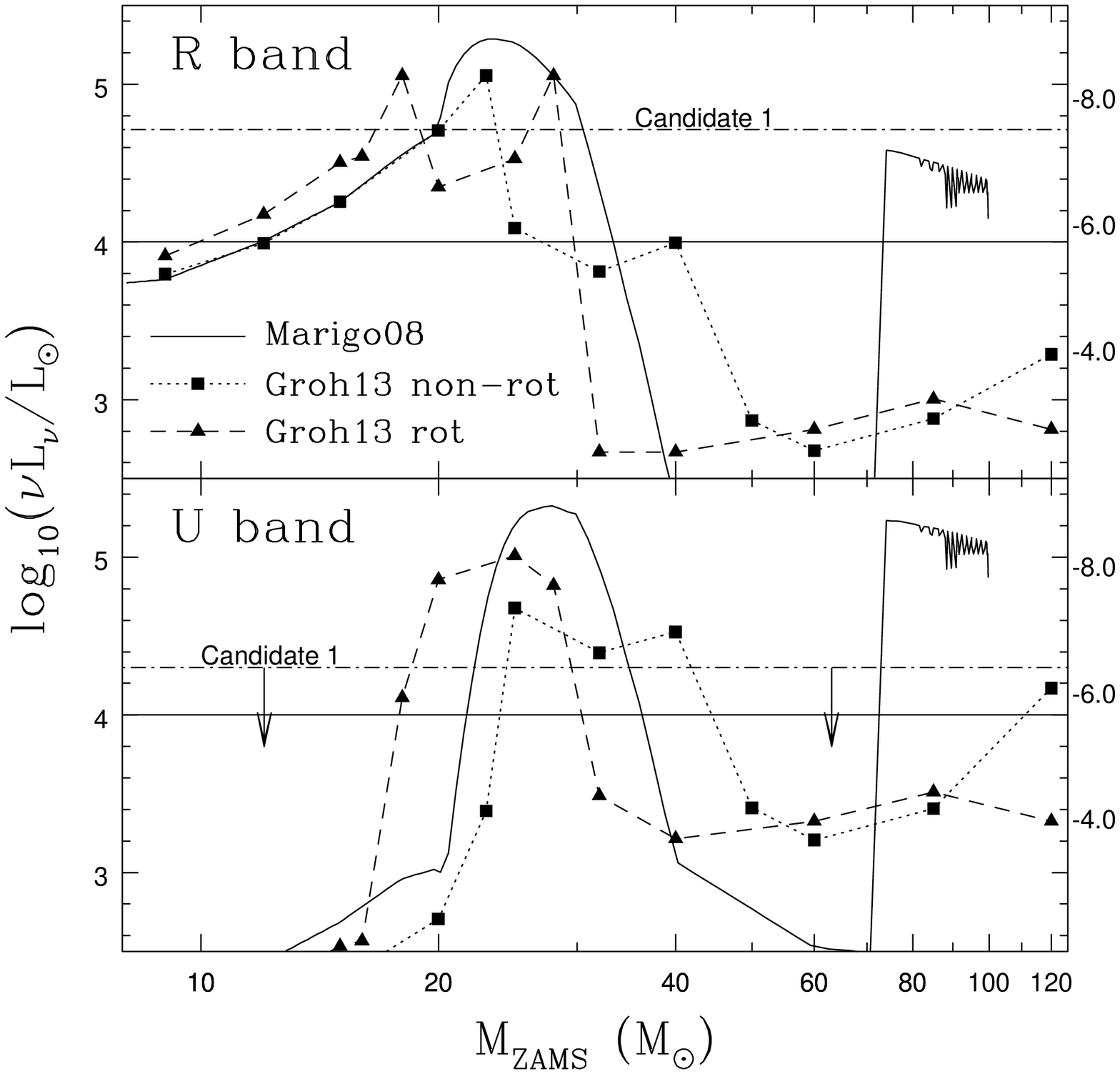}}
  \caption{
    The expected $R_c$ (top) and $U$ (bottom) band luminosities of stars at the end
    of their lives based on \protect\cite{marigo2008} and \protect\cite{groh2013}.  The 
    solid curves are for \protect\cite{marigo2008} and the dotted (dashed) lines are for
    the non-rotating (rotating) \protect\cite{groh2013} models.  The 
    vertical axis is in units of log $(\nu L_{\nu}/L_{\odot})$ on the left and absolute 
    magnitude on the right. The horizontal axis is the 
    zero-age main sequence (ZAMS) mass of a star.  A horizontal line marks 
    $\nu L_{\nu} \simeq 10^4L_{\odot}$, our limit for identifying candidates.
    The luminosities of our remaining candidate, Candidate 1, are indicated by the horizontal 
    dot-dashed lines.
  }
  \label{fig:model2}
\end{figure}

Exposure times were chosen to try to reach a fixed point source luminosity limit for each galaxy, 
although in 
practice, we could not perfectly scale exposure times as the squared distance to 
the galaxies.  
Table \ref{tab:galsam} gives the theoretical depths from the LBC exposure time calculator 
for a signal-to-noise ratio (SNR) of $SNR=5$ for the $R$ band images and $1\arcsec$ seeing. 
These estimates agree well with individual observations ($\sim 0.1$ mag).  
More importantly, we get an average of $\sim 1$ count per $L_{\odot}$ 
($\nu L_{\nu}$ in a band), so the 
massive stars we are interested in should be detectable with enough counts 
to monitor their variability.  
Figure~\ref{fig:model2} shows the expected $R_c$ and $U$ band luminosities of stars at the end
of their lives based on \cite{marigo2008} and \cite{groh2013}.  The primary
differences are due to the effects of mass loss on the effective temperatures.
Most of the mass range of interest can be covered by a search for stars which change 
in luminosity by $\Delta \nu L_\nu > 10^4 L_\odot$.
This will exclude the lowest mass progenitors, $\sim 10M_\odot$, and higher mass
progenitors that have been stripped to become Wolf-Rayet (WR).  
In all these models, however, stars in the $15$-$25M_\odot$ range tend to have
luminosities approaching $\nu L_\nu \sim 10^5 L_\odot$ in at least one of our
survey bands.  Hence we set the sensitivity goal for our survey to reach at
least $\nu L_\nu > 10^4 L_\odot$ to include all the evolved stars 
expected to experience a SN that do not experience 
a high level of mass loss and become the stripped WR stars 
responsible for Type Ib and Ic SNe.

\section{Image Subtraction and Calibration}
The basic data reduction steps of overscan correction, bias
subtraction, and flat fielding are preformed using the IRAF MSCRED package.
While we collect data even in poor conditions on the grounds that an epoch in poor 
conditions is better than no epoch at all, here we only analyze images with 
FWHM $ < 2 \arcsec$.  The median image quality is $1\farcs3$. 
Our search for stars in crowded fields showing large changes in luminosity is 
best accomplished by employing image subtraction.  We use the ISIS image subtraction 
package (\citealt{lupton1998}; \citealt{alard2000}) to process our survey data. 

We median combined a minimum of 3 of the best quality images (e.g. best FWHM, no cirrus, 
no moon) and on average used about 20\% of 
the available images in the reference image.  It is important to use many images to 
construct the reference image to minimize the noise it contributes to the 
subtracted images.  
The image subtraction is preformed by convolving the reference image with a kernel 
so that the psf structures of the observation and the reference image are matched.  
Because the reference image combines the best images, the noise and image quality 
of the subtracted image are dominated by the properties of the image 
being analyzed rather than the reference image.  
Unfortunately the best seeing conditions were rare enough that we cannot, at this point, 
restrict the images used to construct the reference image to the second half of the survey, for example, and maintain 
the necessary quality of the reference image.  

The $R$ band reference image was used as the astrometric reference image for all four bands.  By 
doing this, the $UBV$ and $R$ band images all have the same astrometric solution, and 
it becomes trivial to cross-match sources between bands.  
An astrometric solution was found for the majority 
of the galaxies using the IRAF package MSCTPEAK and SDSS stars \citep{ivezi2007}.  There were 4 galaxies 
(NGC~6946, NGC~6503, NGC~925 and IC~2574) where we could not use 
SDSS to calibrate the reference images.  
Astrometric solutions for NGC~6946, NGC~925 and IC~2574 were found using the USNO-A2.0 
catalog \citep{monet1998}.  
 The astrometric solution for NGC~6503 was found using the HST Guide Star Catalog 2.3 \citep{lasker2008}.
The typical residuals for the astrometric solutions are $\sim0\farcs2$.

Most photometric calibrations were also based on SDSS.
The SDSS photometry was transformed from the SDSS $ugriz$ filter system to the $UBVR_c$ system 
using the prescription described by \citet{jordi2006}.  Thus, the final photometry consists 
of Vega magnitudes with the zeropoints as reported by \citet{blanton2007}.
Again, there were 4 galaxies (NGC~6946, NGC~6503, NGC~925 and IC~2574) where we could not use 
SDSS to photometrically calibrate the reference images. 
The $U$ band photometric solution for NGC~6946 was determined using \citet{botticella2009} 
 for chip 2 and \citet{sahu2006} for chip 1.  The photometric solution for chip 3 
 was determined using overlaps between epochs with chip 2.  The $R_c$ and $V$ solutions 
were found using photometry from \citet{welch2007}.  The $B$ band solutions for chips 2 
and 3 were found using SINGS ancillary data \citep{kennicutt2003}. 
 The $B$ band calibration for chip 1 at NGC~6946 and the $R_c$ band of NGC~6503 were
calibrated using the USNO-B1.0 catalog  \citep{monet2003}.  We required the sources used for 
calibration from the USNO-B1.0 catalog to be detected in both epochs and have the two 
measurements agree to within 0.3 mag. 
 The photometric solutions for the $B$ and $V$ bands of NGC~6503 were found using the 
HST Guide Star Catalog 2.3.
 IC~2574 and NGC~925 were calibrated in the $B$, $V$ and $R_c$ bands using the SINGS ancillary data 
\citep{kennicutt2003}.  For the present study, the $U$ band data for IC~2574, NGC~925 and NGC~6503 
remain uncalibrated.  We are actively addressing calibration issues for future work. 
The typical photometric errors are 0.06 mag.  Since we are looking for large changes in 
luminosity, our absolute photometry does not require a high level of precision, 
although the calibrations of the latter few galaxies clearly require improvement. 
When we select candidates based on luminosity cuts, we correct for Galactic extinction using 
the estimates from \citet{schlafly2011}.  

We mask the images in two different ways.  First, we create a subtraction mask that is applied 
before the ISIS image subtraction.  
On all images we apply a 9 pixel radius mask around any pixel exceeding 60,000 
counts using IRAF.  This removes saturated stars and most of their associated bleed trails, 
which are a dominant cause of erroneous variable sources.  The reference image 
mask combines the masks of all the images used to construct it. This masking significantly 
improves the quality of the image subtraction and reduces the number of spurious 
variable sources.  As discussed below, we catalog and track 
the individual saturated stars that lead to masked regions.  However, we do lose the fainter 
stars that lie in the masked regions.

 Unfortunately, subtractions near the edges of the masked regions are then damaged by the 
 presence of the edge. So for candidate selection we use a ``survey'' mask.  This 
 second type of masking expands 
 the subtraction masks a further 5 pixels and also masks the chip edges where image 
 centering shifts produce subtraction artifacts. 
To estimate the fraction of the galaxy we are surveying, we calculate the masked area.  
Table~\ref{tab:masks} reports the fraction of each pointing/field that is masked for both 
the individual filters and the fraction that is masked in all filters (i.e., the ``logical and'' 
of the masks).  

\begin{table}
\caption{Field Masking Percentage Fractions} 
\label{tab:masks} 
\begin{tabular}{lcccccc}
  \hline \hline
  Galaxy & Chip & U &  B & V &  R & All \\
  \hline 
M81 & 1 & \s4.0 & \s 5.9 & \s 6.8 & \s 7.5 & \s 3.2 \\
 & 2 & \s4.0 & \s 7.3 & \s 9.3 & \s 8.7 & \s 3.2 \\
 & 3 & \s3.4 & \s 7.4 & \s 8.2 & \s 5.5 & \s 2.9 \\
 & 4 & \s4.4 & \s 5.8 & \s 6.9 & \s 6.3 & \s 3.0 \\
M82 & 2 & 10.2 & 13.7 & 15.0 & 11.6 & \s 6.9 \\
M101 & 1 & \s5.0 & \s 5.3 & \s 6.0 & \s 5.1 & \s 3.4 \\
 & 2 & \s4.0 & \s 4.6 & \s 5.4 & \s 5.2 & \s 2.9 \\
 & 3 & \s3.7 & \s 4.2 & \s 5.1 & \s 4.6 & \s 2.7 \\
 & 4 & \s5.0 & \s 5.2 & \s 5.5 & \s 3.4 & \s 2.6 \\
N628 & 1 & 12.7 & \s 8.6 & \s 9.7 & \s 6.7 & \s 3.5 \\
 & 2 & \s9.0 & \s 6.6 & \s 7.7 & \s 7.2 & \s 4.5 \\
 & 3 & \s9.3 & \s 9.8 & \s 7.8 & \s 7.1 & \s 4.0 \\
N672 & 2 & \s5.6 & \s 9.3 & \s 9.9 & 10.0 & \s 4.4 \\
 & 4 & \s6.7 & \s 7.5 & \s 8.9 & 11.2 & \s 4.5 \\
N925 & 2 &        &  \s8.6 & 10.6 & 14.9 &  \s6.3 \\
N2403 & 1 & \s4.8 & \s 5.3 & \s 6.5 & \s 4.3 & \s 3.0 \\
 & 2 & \s4.0 & \s 5.0 & \s 7.9 & \s 7.7 & \s 3.0 \\
 & 3 & \s4.1 & \s 4.7 & \s 5.6 & \s 4.6 & \s 3.0 \\
 & 4 & \s5.3 & \s 7.7 & \s10.0 & \s12.8 & \s 3.0 \\
N2903 & 2 & 13.9 & 15.6 & 14.4 & \s 8.1 & \s 4.5 \\
N3077 & 2 & \s5.9 & \s 8.5 & \s 9.5 & \s 6.7 & \s 3.3 \\
N3344 & 2 & \s6.3 & \s 7.7 & \s 9.9 & 12.6 & \s 5.3 \\
N3489 & 2 & \s4.0 & \s 4.2 & \s 5.3 & 11.1 & \s 3.6 \\
N3627 & 3 & \s5.2 & \s 6.1 & \s 7.3 & \s 9.3 & \s 3.2 \\
N3628 & 1 & \s5.0 & \s 5.1 & \s 9.4 & \s 3.5 & \s 2.8 \\
N4214 & 2 & \s6.4 & \s 6.4 & 10.5 & \s 6.8 & \s 5.1 \\
N4236 & 2 & 11.8 & \s 6.4 & 12.6 & \s 6.8 & \s 5.6 \\
N4248 & 4 & \s6.6 & \s 6.7 & \s 6.7 & \s 5.8 & \s 4.1 \\
N4258 & 2 & \s4.8 & \s 4.6 & \s 5.1 & \s 7.1 & \s 3.0 \\
N4395 & 2 & \s9.0 & \s 9.1 & \s 9.4 & \s 7.3 & \s 6.6 \\
N4449 & 2 & \s5.8 & \s 6.1 & \s 5.6 & \s 5.3 & \s 4.4 \\
N4605 & 2 & \s4.9 & \s 5.2 & \s 8.8 & \s 3.8 & \s 3.1 \\
N4736 & 2 & 12.1 & 12.6 & 12.8 & \s 3.2 & \s 2.5 \\
N4826 & 2 & \s5.0 & \s 4.8 & \s 6.0 & \s 4.4 & \s 3.1 \\
N5194 & 2 & \s6.9 & \s 7.9 & \s 8.2 & \s 6.2 & \s 4.5 \\
N5474 & 2 & \s5.9 & \s 6.2 & \s 7.5 & \s 4.7 & \s 3.0 \\
N6503 & 2 &        & \s 9.7 &  \s7.8 & \s 8.5 & \s 4.7 \\
N6946 & 1 & \s7.4 & \s 9.8 & 17.8 & 19.7 & \s 4.0 \\
 & 2 & \s8.5 & 11.7 & \s22.1 & 25.8 & \s 5.3 \\
 & 3 & \s7.0 & \s 8.5 & 17.8 & 20.3 & \s 4.7 \\
I2574 & 2 &        &  \s7.4 &  \s7.8 &  \s5.2 &  \s4.1 \\
\hline
\multicolumn{7}{p{0.4\textwidth}}{The percentage fraction of each pointing/field 
that is masked both for the individual filters and in all filters (i.e., the ``logical and'' 
of the masks).  The missing $U$ band entries are those lacking a $U$ band photometric calibration.}
\end{tabular}
\end{table}

\section{Candidate Selection}
To search for variable and ultimately vanishing stars, we combine several approaches to target 
selection.  We start with three lists of potential candidates.  We perform PSF photometry using 
DAOPHOT on the reference image and create a list of bright sources.  We define a bright source 
as one having an observed luminosity of $\nu L_{\nu}  \ge 10^4L_{\odot}$ in any band.  
Second, we identify all the variable sources in our survey.  The variable sources are identified 
by performing aperture photometry using SExtractor on the RMS image.  The RMS image is 
the pixel by pixel root mean square (rms) average of the subtracted images formed after eliminating 
those with seeing FWHM above $2\farcs0$.  Each subtracted image is convolved with a Gaussian 
of width, $\sigma^2= (2\farcs0^2-FHWM^2)/8\ln(2)$, designed to give all the subtracted images 
a similar resolution of $2\farcs0$.  Finally, we created a list of saturated sources that are masked 
in the reference image.  

To remove artifacts from our source lists, we impose a cut on the flux ratio of two different 
apertures on the RMS image.  Effectively, this cut removes SExtractor detections that are not 
point-like variable sources. 
We preformed IRAF aperture photometry on the RMS image using a 4 and an 8 pixel 
radius aperture, $F_{4}$ and $F_{8}$, for both our variable and bright sources. 
Experiments and visual inspection demonstrated that $F_{4} / F_{8} > 0.4$ for real sources, and we 
reject a source that fails this test in all four filters.  

We then combine the bright and variable lists, matching sources that are within 1 pixel of each 
other. Finally, we add the list of masked stars to create a final list of targets.  ISIS light curves 
are then created for all targets that are not masked in the reference frame.  These target lists 
are independently created for each of the 4 bands.  For candidate selection we examine 
the light curves from the beginning of the survey through the observing run ending 
10 January 2013.  The first and 
last observation dates for each galaxy are reported in Table~\ref{tab:galsam}.
We use the data taken after the January~2013 run
to help determine the nature of any candidates.

Ideally, we would have an accurate magnitude for each source in the reference image which we 
would use to normalize the difference imaging light curves.  We would then simply analyze 
these light curves.  In practice, confusion and saturation in the reference image means that we 
must consider several different cases.  
 Furthermore, whether a target is masked can change over the course of the survey.  Some of 
 the changes are caused by astrophysical phenomena, like novae.  Some of the changes are 
 caused by properties of the data such as seeing variations or the size and rotation of the 
 diffraction patterns from bright stars.  
We can broadly divide the possible scenarios by whether the source is (1) never masked, 
(2) masked in the reference image, or (3) unmasked in the reference image but masked 
in some observational epochs.

\subsection{Sources Unmasked in the Reference Images}
The targets found on the list of bright and/or variable sources are by definition unmasked in 
the reference image.  
For each source we produce an unnormalized light curve in 
$\Delta \nu L_{\nu}(t)=\nu L_{\nu}(t)- \nu L_{\nu}(ref). $  We would also like an estimate of 
the normalized light curve, $\nu L_{\nu}(t)$, which requires an estimate of the source 
luminosity, $\nu L_{\nu}(ref)$, in the reference image.  While $\Delta \nu L_{\nu}(t)$ suffers 
little from crowding or blending, there are challenges for estimating $\nu L_{\nu}(ref)$.

We estimate $\nu L_{\nu}(ref)$ in one of three ways.  If a target matches a source in our full 
DAOPHOT catalog for the reference image, then we use this to normalize the light curve.  
While all bright targets have a DAOPHOT magnitude by definition, not all of the variable 
targets will be found as a source in the reference image.  If no DAOPHOT source is found, we 
perform aperture photometry using IRAF on the reference image at the target location on the 
RMS image.  If a significant flux is detected, we use this to normalize the light curve.  Aperture 
photometry typically failed in areas of high background or near masked pixels.  
If aperture photometry fails, the light curve is normalized to the largest decrease in luminosity 
seen in the light curve since the target must be at least that bright.  Essentially, we set the 
minimum $\Delta \nu L_{\nu}(t)=0$.

Given $\Delta \nu L_{\nu}(t)$ and our best estimate of $\nu L_{\nu}(t)$ we define several broad 
criteria to select candidates for failed SNe.
The first set of criteria performs a general search for a source that has a large decrease in 
luminosity, possibly vanishing by our last observation. These criteria make no assumptions 
about the nature of any transient associated with black hole formation.  We are simply looking
for a vanishing massive star.  
As discussed earlier, the Padua stellar models \citep{marigo2008} or the \citet{groh2013} 
stellar models 
show that a luminosity limit of $\nu L_{\nu} \sim 10^4L_{\odot}$ will capture the evolved stars 
expected to experience a SN that do not experience 
a high level of mass loss and become the stripped Wolf-Rayet stars responsible for Type Ib and Ic SN 
(see Figure \ref{fig:model2}).
We simply calculate the change in luminosity between the first and last observation from the 
differential light curves. If the change in luminosity is 
$\nu L_{\nu}(t_1)-\nu L_{\nu}(t_N) \ge 10^4L_{\odot}$, the source is considered a possible 
candidate.  Note that this criterion is independent of $\nu L_{\nu}(ref)$.
We also identify objects that become significantly brighter 
$\nu L_{\nu}(t_N)-\nu L_{\nu}(t_1) \ge 10^4L_{\odot}$ and follow these sources as we would a 
fading possible candidate.  This criteria was designed to be broad, not requiring any particular 
signature for a failed SN since the optical signature is uncertain.  
Instead we search for a source that is clearly detected at the beginning of the survey and at some 
point becomes undetectable for the remainder of the survey.  No matter what the signature of a failed 
SN, the progenitor star will ultimately vanish in our bands.  
The brightening candidates allow us to explore our false-positive 
rate and will provide a sample of novae and other bright variables. 

The second set of criteria address sources that become masked at some point in the survey, but 
are not masked in the reference image.  Changes in a target's masked status can be due to 
astrophysical phenomena as well as changes in the data, such as small changes in the rotator 
angle.  It is very difficult to automatically separate these two causes for a masking change. 
However,  our multiple bands help minimize this problem.  For example, if a target in the $R_c$ band 
moves in and out of our masking limit with changing seeing, it is very likely this source will be 
unmasked in one of the bluer bands. 
Since we do the initial candidate selection independently in each band, sources that are 
occasionally masked in a band are considered a possible candidate if the target is masked at 
any point in the survey and is found during our last observation to have a luminosity 
$\nu L_{\nu} \le 10^4L_{\odot}$.  When we match candidates between the bands, as will be 
described later, we examine such sources in bands where they are unmasked, if possible.

Our last set of criteria is designed to identify the failed SN signature identified 
by \citet{lovegrove2013} for red supergiant progenitors. While the shock breakout 
during a failed SN, discussed in \citet{piro2013}, produces a brighter optical transient, 
it only lasts for 3--10 days and this is too short to search for given our survey cadence. 
\citet{lovegrove2013} found that for $15-25M_{\odot}$ 
red supergiants the envelope of the stars becomes unbound, producing a transient that lasts 
about $\sim1$~year with a luminosity of order $L_{bol} \sim 10^6L_{\odot}$ and a temperature 
of $\sim4000$ K at peak.  We can easily search for such a source by selecting all 
targets that have a luminosity of $\nu L_{\nu} \ge 10^5L_{\odot}$ for between 3 months and 
3 years during the survey.  We impose this broad timing window to exclude objects that are 
always high luminosity or reach this luminosity only briefly.  The luminosity limit is lower 
than $L_{bol} \sim 10^6L_{\odot}$, because we must take into account bolometric corrections.  The changes in the 
blue bands will be smaller given the expected temperature, so this is primarily a search in 
the $R_c$ and $V$ bands. 
 
\subsection{Saturated Stars in the Reference Images}
The reference images are constructed from the best quality images of 
each galaxy.  This means that the reference image is generally comprised of 
images with the smallest FWHM, leading to the maximum number of saturated stars.  
The absolute magnitude at saturation varies from galaxy to galaxy, ranging from roughly $-9.5$ to $-11.5$ mag 
in the $R_c$ band and from roughly $-10$ to $-12.5$ mag in the $U$ band.   
The sources 
masked in the reference image are not necessarily masked in all the survey images, either 
because of changes in seeing or due to actual luminosity changes of the source.
For example, a star that is saturated in $R_c$ may not be saturated in $U$ band.
  For saturated 
stars in the reference image we simply check their fluxes in the final epoch.  We search the DAOPHOT 
catalog of the last image within a 3 pixel radius around the location of the star and pick the 
brightest source as our match.  We search a larger radius because the location determined for 
the masked star in the reference image is approximate.  If the source is still saturated or bright 
we ignore it.  However if the luminosity in the final image is $\nu L_{\nu} \le 10^5L_{\odot}$, the 
source is considered a possible candidate and is visually inspected.  If no match is found, aperture 
photometry is preformed at the location.  If the source is still masked, it is considered saturated 
and not a candidate.  If a non-zero flux is found, the source is considered a possible candidate. 

\begin{figure*}
\centerline{\includegraphics[width=7.0in]{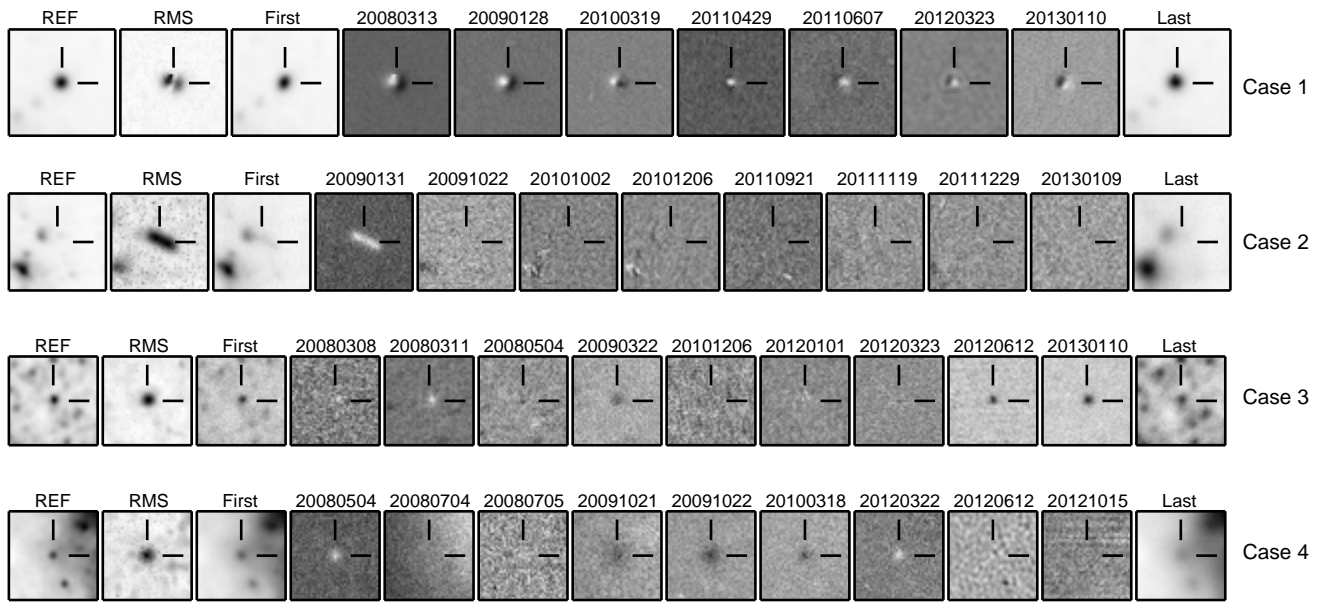}}
\caption{
Four examples of sources that were selected as candidates 
and removed from the candidate list through visual inspection.  
In the subtracted images, which show the individual images minus the 
reference image, darker shades mean the source is dimmer than in the 
reference image and whiter shades mean it is brighter.  For the REF, RMS, First, 
and Last images, the darker the source, the brighter it is.  
Case 1 is a bright star with ``dipole" subtraction residuals.  These are the most 
common false positives.  
Case 2 is an asteroid.  These are relatively uncommon and are easily removed 
due to their motion.
Case 3 and Case 4 are both variable stars that happen to be dim at the end of 
our selection period. In most cases they are easily removed because they 
continue to vary after the end of the survey period.  
We show the reference image, the RMS image (where the brighter the source, the more it varied over the 
course of the survey), the first observation, a selection of the 
subtracted images labeled by their epoch and, finally, the last 
observation.  The images are 10 arcseconds on a side.
}
\label{fig:ex_cut}
\end{figure*}

\subsection{Summary of Candidate Identification}
To summarize, the initial target list was comprised of sources that were not identified as 
artifacts and satisfied one of these three criteria.

\begin{enumerate}
\renewcommand{\theenumi}{(\arabic{enumi})}
\item Luminous ($\nu L_\nu > 10^4 L_\odot$) in the reference image, 
\item A source in the RMS variability image, or
\item Masked at some point, 
\end{enumerate}

\noindent in any of the four bands.  
Each such target was then considered a candidate if it showed either 
\begin{enumerate}
\renewcommand{\theenumi}{(\arabic{enumi})}
\item A difference between the first and last observation of $ | \Delta \nu L_\nu | > 10^4 L_\odot$,
\item $\Delta \nu L_\nu > 10^5 L_\odot$ for 3 months to 3 years during any point of the survey, or
\item It was masked at some point in the survey but was found to be unmasked and no longer 
bright in the last observation.
\end{enumerate}

\subsection{Processing the Candidates}
Once we identify possible candidates in each of the bands, we cross match the bands using a 
1 pixel radius.  We next compare the DAOPHOT PSF luminosity from the last observation to 
that found by ISIS for each band in which the source is a candidate.  If in any band these 
measurements match to within 0.3 mag and the candidate has a luminosity in the final image 
above $\nu L_{\nu} \ge 0.25 \times 10^4L_{\odot}$, then the detection and measurement are 
considered secure, and we eliminate the candidate.  If the photometry in the final 
image does not meet these criteria, then the candidate is visually inspected. 

At this point we were left with 11,134 candidates across all galaxies and filters.
Two of the authors (JG and CSK) visually inspected all the remaining candidates.  
We inspected candidates that are never masked in the longest wavelength band they are 
found as a candidate.
If a candidate is sometimes masked or masked in the reference image, and is never 
masked in another band, we inspected it in that band.  If the candidate is unmasked in multiple 
bands, we inspected it at the longest wavelength band available. 
Often, sources masked in the $R_c$ band are not masked in the $U$ band.  If the target is sometimes 
masked in all bands, we inspected the target in the longest wavelength band it was found as a 
candidate. 

For the visual inspection, we looked at postage stamps around the source in the reference 
image, the last observation, the RMS image and each subtracted image in the survey.
We constructed light curves and selected candidates from the survey data taken through 
January 2013.  Data taken after this period were used to help determine if the source 
remained faint or was a persistently variable source. 
If we detect the source after January 2013 either directly in the image or through variability 
we declare that the source is not a failed SN and no longer a candidate. 

The most common classifications of sources that did not survive as candidates are
true variable stars that are clearly present after January 2013 and 
inconsistent subtraction of bright stars creating false candidates.  
Figure~\ref{fig:ex_cut} shows 4 different examples of sources that were selected as 
candidates and removed through visual inspection.  Case 1 (top) shows an example 
from NGC~5474 of the inconsistent subtraction of bright stars.  Such sources are 
easy to identify and remove from the candidate list.  Case 2 shows an 
asteroid in the field of NGC~628.  
Case 3 and Case 4 both show variable stars.  Case 3 is in M101 and fades over the course 
of the survey selection period, but is 
clearly present in the last observation.  Case 4 shows a variable star in NGC~6503 that 
does not steadily fade.  It happens to be dimmer at the end of our candidate selection 
period in 2013, but the bright star is still visible in the last observation from 2014.  
This case also 
illustrates how poorer quality data, like the last observation NGC~6503, causes ambiguity. 
Many of the candidates that survived the first round of inspections were from galaxies that 
were less well sampled or had poorer quality data.  
Candidates were kept if either inspector considered it a candidate. 

This first round of visual inspection resulted in a list of 235 
residual candidates.
  Candidates that passed through the first round of visual inspections 
were then examined a second time using all available bands,
which helps clear up most remaining ambiguities.  For example, an area 
which is crowded in the $R_c$ band may be much more sparse in the $U$ band and something near 
the detection threshold in the $V$ band may be a clear detection in the $R_c$ band. 
This more detailed inspection of the candidates reduced the number of candidates to 14.  
A final round of inspections with additional data was completed adding the calibrated light 
curves to both 
aide our interpretation of the subtracted images and for detailed checks of their light 
curves.  We were left with 4 final candidates, which we will discuss in more detail 
in sections 6 and 7. 

In addition to selecting our failed SN candidates, our search 
allows us to compare the number of variable sources that had faded at the 
end of the selection period to those that had brightened.  We would expect there to 
be about the same number of sources that had increased in brightness as those that 
had decreased in brightness and indeed this is the case.  We find 3586 fading variable 
sources and 3514 brightening variable sources, giving a ratio of $1.02 \pm0.02$.  
This helps show that we 
are not preferentially detecting sources that fade, but instead detect variable sources 
that are fading and brightening with roughly equal efficiency.  We also detected 
$\sim 40$ sources that met our Lovegrove-Woosley model based criteria, 
but all of these sources were ultimately rejected. 
 
\begin{figure}
\centerline{\includegraphics[width=3.5in]{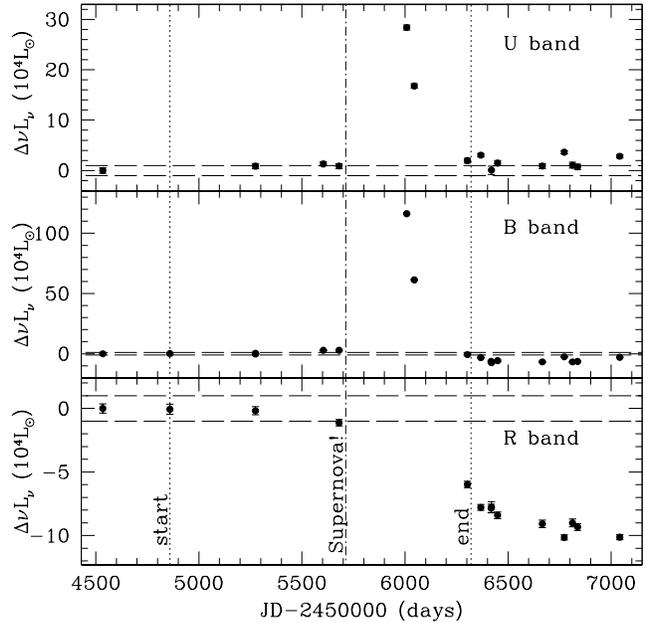}}
\caption{The $U$ (top), $B$ (middle) and $R_c$ (bottom) band differential light curves 
for SN~2011dh in M51.  For the $U$ and $B$ bands we show the results
as originally processed, while for $R_c$ we show a reprocessed series where the reference image included 
only images prior to the SN.  The vertical axis is in units of $10^4L_{\odot} (\nu L_{\nu})$ and has 
been normalized to the first observation so that the luminosity difference between the first and last observations can be 
easily seen.  Note, however, the change in luminosity scale between the $U/B$ images showing the 
SN and the $R_c$ images that do not.  
A change in luminosity by $10^4L_{\odot}$ in either direction, marked by the horizontal lines, 
would lead to the source being selected as a candidate.  The vertical dotted lines show the 
beginning and end of the candidate selection period.
}
\label{fig:2011dh_lc}
\end{figure}
\begin{figure*}
\centerline{\includegraphics[width=7.0in]{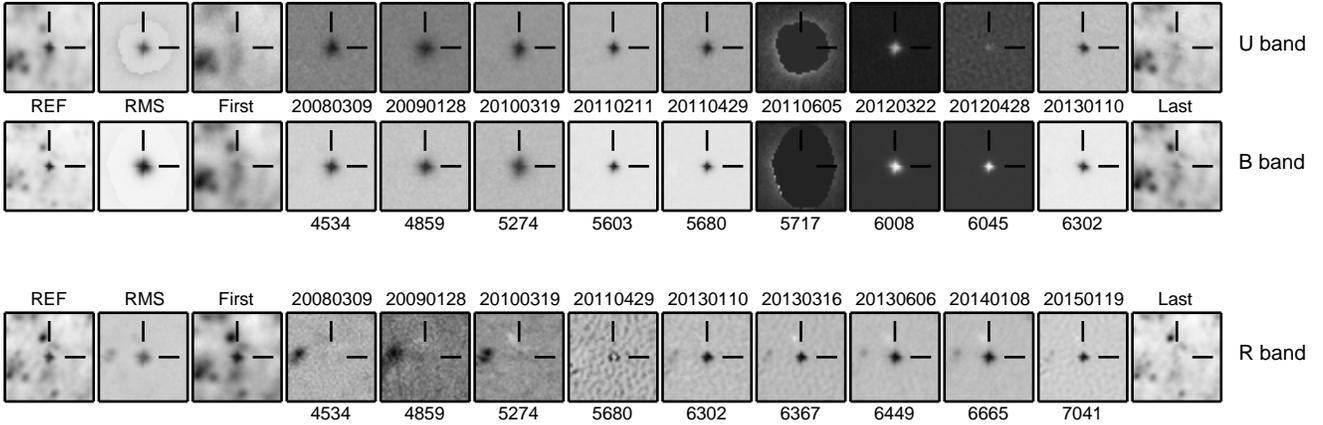}}
\caption{The top and middle rows show the $U$ and $B$ band images for SN~2011dh from our original candidate 
selection process and the bottom shows the $R_c$ band images from our experiment creating 
a reference image using only observations taken prior to the SN.  We show $R_c$ band images 
without the SN from the candidate selection period and only 
select observations from afterward. 
The ``First'' observations are on 9 March 2008 and the ``Last'' observations shown in the 
$U$ and $B$ bands are on 10 January 2013 and on 19 January 2015 for the $R_c$ band. 
The format is the same as in Figure~\protect\ref{fig:ex_cut}.
}
\label{fig:2011dh_image}
\end{figure*}

\section{Successful Supernovae}
We can use the successful core-collapse SNe (ccSNe) in these galaxies as tests of our approach.  
We processed these sources just as we would any star, basically preforming a blind search 
for a vanishing star, as long as our last observation is 
taken after the SN has sufficiently faded.  During our survey period there were 
three core-collapse SN (SN~2009hd, SN~2011dh, and SN~2012fh) 
and one Type Ia SN (SN~2011fe) in our sample.  
Here we discuss the 3 ccSNe that occurred during our search window. 

The first SN that occurred during our candidate selection period was
SN~2009hd in NGC~3627 (\citealt{monard2009}; \citealt{elias2011}).  This SN exploded 
in June 2009 when we had 3 epochs of data.  
The SN occurred in a dusty region and because of the resulting low progenitor fluxes
and the sparsity of epochs we have no useful information on the variability of the progenitor. 
We have 4 epochs after the explosion and 
during our candidate selection period, with the last observation on 28 April 2012, plus 5 
additional epochs after the selection period. 
This source was chosen as a brightening candidate 
in the $U$, $B$, and $V$ bands.  In the $R_c$ band, the source fell next to a bright star that is masked in our 
images. 
The luminosity of the SN is still brighter than the progenitor even in 
2014, probably because of interactions between the SN shock and the circumstellar medium. 
As a result, we did not find it as 
a faded star.  A failed SN, even with mass ejection as in \citet{lovegrove2013},
could not produce this bright long-term luminosity because there would be either no shock or 
a very weak shock.  
Shock driven luminosities scale as $L \propto v_{s}^3$ where $v_{s}$ is the shock speed, so a failed SN 
with $v_{s} \sim 200$ km/s would be about $(200/4000)^3 \sim 10^{-4}$ times less luminous than 
a true SN with  
$v_{s} \sim$ 4000 km/s and could not sustain the luminosity of a massive evolved star.

The best test of our search algorithm is SN~2011dh in NGC~5194. This type IIb SN 
was discovered quickly after its May 31 explosion \citep{griga2011}.
At the time of the SN, we had 5 epochs in UBV and 4 epochs in $R_c$.  The progenitor was 
easily visible and found to be variable in our LBT data, as discussed in detail in 
\citet{dorota2012a}.  We have 4 additional epochs during our selection period 
after the SN occurred.  The last observation of the source in our 
candidate window was on 10 January 2013, one of the last observations to be included.  

SN~2011dh was selected as a candidate in all four bands, although not truly based
on the star vanishing.  In the present data, the source was selected because of the SN 
surviving as a candidate in the $U$, $V$ and $R_c$ bands.  In the $U$ and $B$ bands, the SN exceeded the
``brightening'' luminosity  ($\Delta \nu L_{\nu} \ge 10^4L_{\odot}$) and it was kept as a candidate in $U$ both for brightening 
and matching the Lovegrove-Woosley model based criteria.  It was not kept as a 
candidate in the $B$ band, where it had not faded sufficiently.  It would have been selected 
at $B$ band had we used the next epoch in March 2013.  
In the $R_c$ and $V$ bands it was selected because it had been masked in the reference image 
due to the inclusion of the SN and 
was unmasked and not found by DAOPHOT in the final epoch ($V$ band) or found to have 
significantly faded ($R_c$ band).  The inspections periodically led to a 
frisson of excitement when this source came up because the vanishing of the progenitor is 
unambiguous in any visual inspection of the later data. 

As an experiment, we explored what would happen if the SN was removed from the $R_c$ band 
data.  We 
created a new reference image using images from before the SN and removed observations where the 
SN was masked.  Figure \ref{fig:2011dh_lc} shows the differential flux light curves for the $U$ 
and $B$ bands from our original blind analysis while the $R_c$ band light curve is from our analysis 
removing the SN.  
The progenitor began with a luminosity of $8.3 \times 10^4L_{\odot}$ in the $U$ band, $2.01 \times 10^5L_{\odot}$ in the $B$ band 
and $1.08 \times 10^5L_{\odot}$ in the $R_c$ band.  The $R_c$ band source is clearly fading and the source
would have been considered a fading candidate during the candidate selection period. 
Figure \ref{fig:2011dh_image} shows the $U$ and $B$ band images from our original candidate 
selection and select $R_c$ band images from all survey observations. 
We show the RMS image (where the brighter the source, the more it varied over the 
course of the survey), the reference image, the first observation, the 
subtracted images labeled with the epoch and finally the last 
observation. In the $U$ and $B$ bands there was not enough time to fade below the luminosity of
the progenitor.  In the $R_c$ band, the subtracted images show the low level of fading found 
by \citet{dorota2012a} before the SN.  There is a definite signal afterwards, 
signifying the lack of flux from the progenitor.  If we analyze this source, it is flagged as a 
vanishing star candidate.  If we include the data later than January 2013 the signal becomes even 
stronger, as shown in Figure \ref{fig:2011dh_lc}.  This shows that our methods are valid and able to discover 
vanishing stars. 

The final ccSN that occurred during our candidate selection was the Type Ic SN~2012fh in 
NGC~3344 \citep{nakano2012}. 
While there are 8 epochs of observations between 04 May 2008 and 23 March 2012, there are
no observations between when the SN occurred (likely 2013/06, \citealt{nakano2012})
and 10 January 2013, the end of our candidate selection period.  Therefore, we cannot use this 
SN to test our procedures.  

\begin{figure}
\centerline{\includegraphics[width=3.5in]{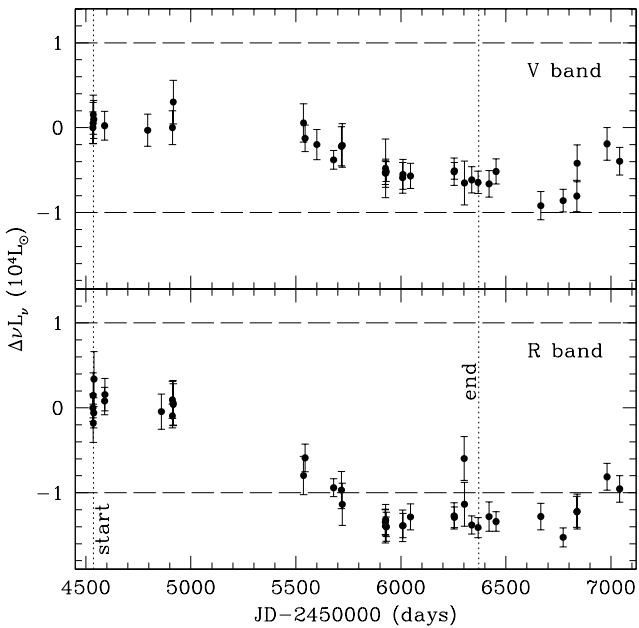}}
\caption{The $V$ (top) and $R_c$ (bottom) band differential light curves for Candidate 2 in 
NGC~4248.  The vertical axis is in units of $10^4L_{\odot} (\nu L_{\nu})$ and has 
been normalized to the first observation so that the luminosity difference between 
the first and last observations can be easily seen.  A change in luminosity by 
$10^4L_{\odot}$ in either direction, as indicated by the horizontal lines, 
would lead to the source being selected as a candidate.
}
\label{fig:4258can_lc}
\end{figure}
\begin{figure*}
\centerline{\includegraphics[width=7.0in]{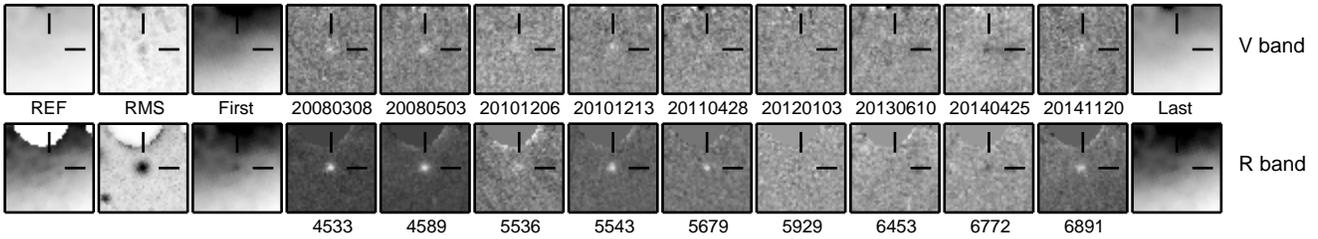}}
\caption{Selected $V$ and $R_c$ band observations of Candidate 2 in 
NGC~4248.  We have 29 epochs for this galaxy in total.  
The ``First'' observation is on 8 March 2008 and the ``Last'' observation 
is on 20 November 2014.  
The format is the same as in Figure~\protect\ref{fig:ex_cut}.
}
\label{fig:4258can_image}
\end{figure*}

There are 2 remaining events that are special cases. 
We caught the transient SN~2008S in NGC~6946 in outburst at the beginning of our survey. 
We do not have any pre-explosion images from this survey, as our first observation was 
taken 3 May 2008. Although \citet{prieto2008} discusses the progenitor of SN~2008S using 
LBT observations taken before before the explosion, this data was not included in our 
analysis.  In our survey, this source was 
chosen as a candidate in all 4 filters due to a decrease in luminosity 
of $\nu L_{\nu}\ge 10^4$ between the first and last observations.  
SN~2011fe, a Type Ia in M101 \citep{nugent2011}, occurred in our sample during our 
candidate selection period.  This source was selected as a candidate 
due to the SN explosion itself. It had not faded enough by the end of our selection range 
and it will not test our methods because the progenitor is constrained to be far 
fainter than an evolved massive star \citep{li2011}.  

\section{Ultimately Rejected Candidates}
Our survey produced 4 final failed SN 
candidates, two of which were promising, Candidates 1 and 2, and two which were more ambiguous, Candidates 3 and 4.  
With additional LBT data from late 2014 and early 2015 we can reject Candidates 2, 3 and 4 with reasonable confidence, 
with additional evidence from archival $HST$ observations to support the rejection of Candidate 3. 
This leaves Candidate 1 as the only current candidate. 
We discuss the ultimately rejected Candidates 2, 3 and 4 to illustrate the selection process. 

These three sources all showed decreases in luminosity between the first and last 
observation of $\nu L_{\nu} \ge 10^4L_{\odot}$.
Candidate 2 was considered a good candidate, with a clearly visible source at 
the start of the survey that faded to become undetected.  Candidates 3 and 4 
were more ambiguous, as they are dimmer sources next to brighter 
stars in galaxies with poorer data.  One candidate is next to a masked region 
of a bright star and the other is a red source that is blended with a 
brighter star.  
Here we explore why these 3 candidates were selected, then removed, from candidacy. 

For a vanished star, the change in luminosity we measure between our first and 
last observations during our candidate selection period is exactly equal to the 
total luminosity of the star if it has vanished. 
Since two of our sources are in crowded regions, these differential magnitudes will be 
a more accurate measure of the SED of a star which has actually vanished 
than direct photometry on the initial observations.

\begin{figure}
\centerline{\includegraphics[width=3.5in]{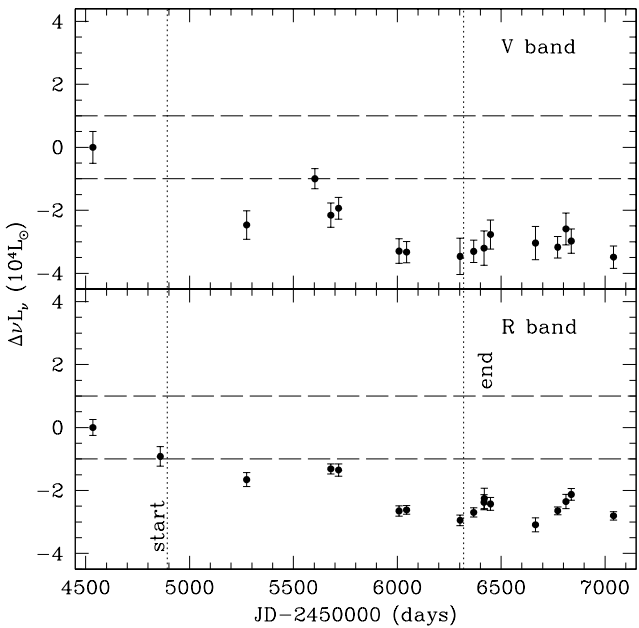}}
\caption{The $V$ (top) and $R_c$ (bottom) band differential light curves for Candidate 3 in 
NGC~5194.  The star is on the edge of a masked region, leading to poorer than usual photometry. 
The format is the same as in Figure~\protect\ref{fig:4258can_lc}.  
}
\label{fig:5194can_lc}
\end{figure}
\begin{figure*}
\centerline{\includegraphics[width=7.0in]{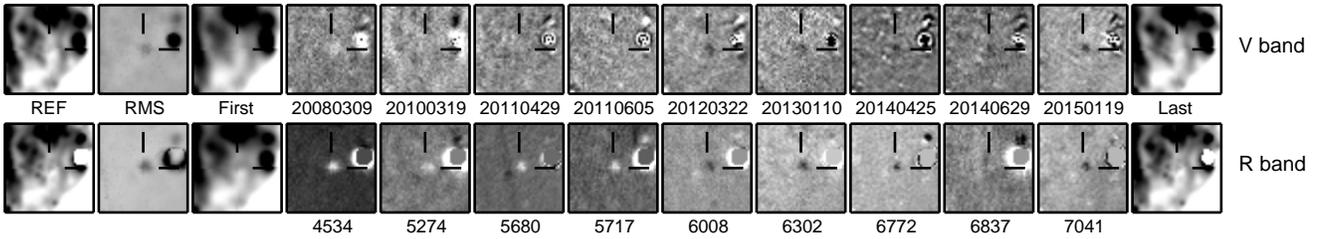}}
\caption{Selected $V$ and $R_c$ band observations for Candidate 3 in 
NGC~5194.  The ``First'' observation is on 9 March 2008 and the ``Last'' observation 
shown is from 19 January 2015.  The format is the same as in Figure~\protect\ref{fig:ex_cut}.
}
\label{fig:5194can_image}
\end{figure*}
\begin{figure}
\centerline{\includegraphics[width=3.5in]{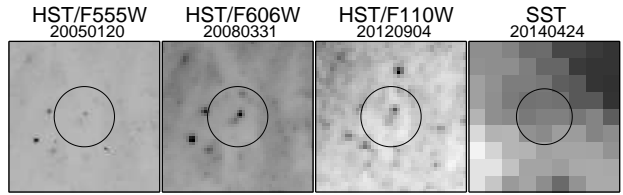}}
\caption{Selected archival observations for Candidate 3 in NGC~5194.  From left to right the 
panels show the $F555W$ {\it HST} observation from 2005 January 04, 
the $F606W$ observation from 2008 March 31, the $F110W$ observation from 2012 September 4
 and an example 3.6$\micron$ {\it SST} observation.  We can see in the {\it SST} 
image that there is mid-IR 
emission in the region, but cannot resolve a particular source. The archival images are 
labeled with the date of the observation.  The circle marks the Candidate 
location and has a 1 arcsecond radius.  The images are 5 arcseconds on a side.  
}
\label{fig:5194_arch}
\end{figure}

\subsection{Former Candidate 2}
This candidate was found in NGC~4248, a dwarf companion galaxy to 
NGC~4258, at RA 12:17:50.56 and Dec +47:24:27.65.  
It was selected as a fading candidate in the $R_c$ band.  
The source was also detected in the $V$ band but not in the $U$ or $B$ bands. Figure 
\ref{fig:4258can_lc} shows the $R_c$ and $V$ band differential light curves for this candidate.  We 
see in both bands that the source begins the survey at a relatively constant luminosity 
for a year.  
We measured the luminosity decrease during our survey period to be 
$1.3 \times 10^4 L_{\odot}$ in the $R_c$ band and find an apparent magnitude 
of $R_c$ $\simeq $ 23.06 mag.  In the $V$ band 
we measure a luminosity decrease of $0.7 \times 10^4 L_{\odot}$ and an apparent 
magnitude of  $V$ $\simeq $ 24.19, giving a $V-R$ color of $\simeq 1.13$ mag.  
Figure \ref{fig:4258can_image} shows select observations from the 28 total epochs for 
this galaxy.  The candidate is no longer visible after 28 April 2011, 
about when the luminosity decrease crossed our threshold to 
consider the source a candidate in the $R_c$ band.
Observations taken with the LBT on 20 November 2014 and 19 January 2015 show the source, 
after almost 3 years below our detection limit, became visible again.  This reappearance 
means this variable star is not a failed SN.  

While there are no {\it HST} archival observations of Candidate 2, there are IRAC
{\it SST} observations from 2007 (Program: 40204, PI: Kennicutt) 
and 2010 (Program: 61008, PI: Freedman).  
There is no source visible at the location of Candidate 2 in either of these epochs.  
For the $3.6\micron$ ($4.5\micron$) data we calculate a $3\sigma$ flux limits of $0.0411\pm0.0005$ mJy 
($0.029\pm0.001$ mJy), which are too weak to constrain the mid-IR luminosity.  

\begin{figure}
\centerline{\includegraphics[width=3.5in]{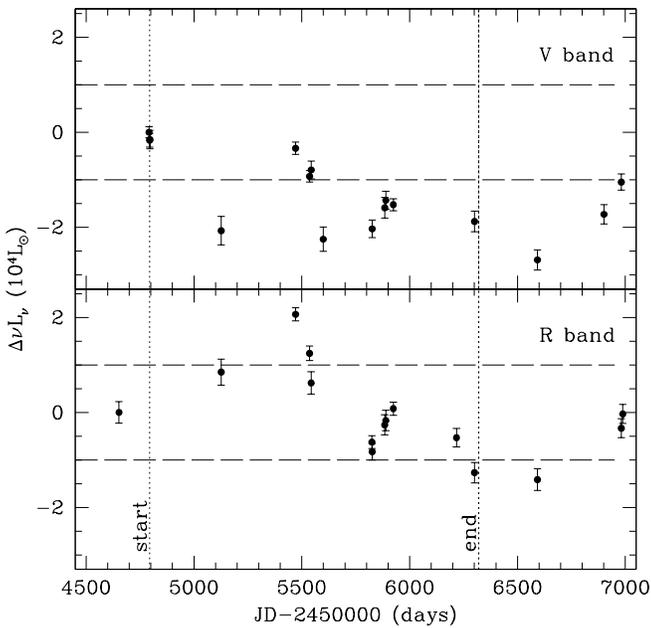}}
\caption{The $V$ (top) and $R_c$ (bottom) differential light curves for Candidate 4 in 
NGC~672.  The format is the same as in Figure~\protect\ref{fig:4258can_lc}.
}
\label{fig:672can_lc}
\end{figure}
\begin{figure*}
\centerline{\includegraphics[width=7.0in]{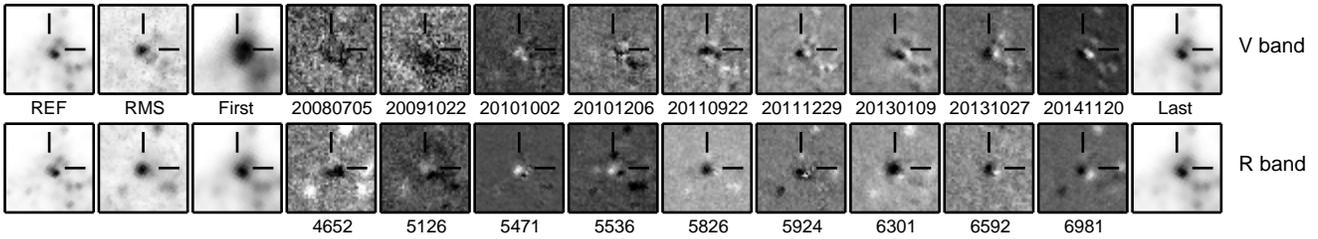}}
\caption{Selected $V$ and $R_c$ band observations for Candidate 4 in 
NGC~672.  The ``First'' observation is on 5 July 2008 and the ``Last'' observation 
shown is from 20 November 2014.  The format is the same as in Figure~\protect\ref{fig:ex_cut}.
}
\label{fig:672can_image}
\end{figure*}
\begin{figure}
\centerline{\includegraphics[width=3.5in]{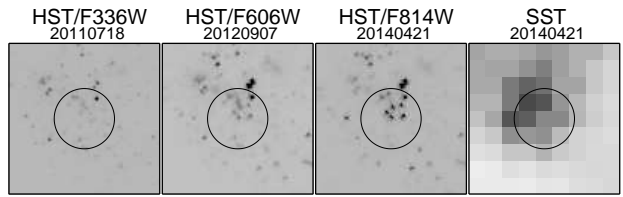}}
\caption{Archival observations Candidate 4.  This figure shows the $F336W$ {\it HST} observation 
from 2011 July 18, the $F606W$ and $F814W$ {\it HST} observations from 2012 September 9 and the 3.6$\micron$ 
{\it SST} observation.  The archival images are labeled with the date of the observation.  
The circle marks the Candidate location and has a 1 arcsecond radius.  The images 
are 5 arcseconds on a side. 
}
\label{fig:672_arch}
\end{figure}

\subsection{Former Candidate 3}
The candidate found in NGC~5194 at RA 13:29:51.01 and Dec +47:11:26.85 was difficult 
to classify given its location and magnitude.  It fell on the edge of a masked region caused 
by a bright star in both the $V$ and $R_c$ bands.  The source 
was selected as a candidate due to fading in the $R_c$ band.  In addition to being next to a 
masked bright star, this 
candidate was also difficult to characterize because it is relatively faint.  We reduced the amount of 
masking used for a nearby bright star to improve the results. The source had $R_c$ $\simeq $ 22.66 mag
with a luminosity decrease of $\nu L_{\nu} = 2.6 \times 10^4 L_{\odot}$.  We find a 
luminosity decrease of 
$3.1 \times 10^4 L_{\odot}$ in the $V$ band and $V$ $\simeq$ 22.81 mag. 
The candidate was not detected in the $U$ and $B$ bands and no variability was seen in those bands.  
Figure \ref{fig:5194can_lc} shows the $V$ and $R_c$ band light curves for this candidate.

The $R_c$ band light curve shows a brightening near the end of the survey that appears to be 
caused by contamination from the neighboring star due to poorer observing conditions.  
The brightening in the $V$ band during the last epoch, which was taken during good 
conditions, is also likely caused by edge effects from the masked star.  No brightening is 
obvious in the raw images. 
Figure \ref{fig:5194can_image} shows selected observations of this candidate. 
This source seems to have increased it brightness on 29 June 2014 and decreased again 
in the last observation on 19 January 2015.  This clear variability is cause to remove this as a candidate.  
The archival $HST$ observations also support eliminating this candidate.  

There are archival data from both {\it HST} and {\it SST} at multiple epochs. 
The {\it HST} observations are available for many epochs and filters, so we only discuss 
select observations from the more standard, wide-field filters.  
We correct the astrometric solution 
to match our LBT reference image and find Candidate 3 corresponds 
to a pair of sources in the {\it HST} observations.  Figure~\ref{fig:5194_arch} shows 
example {\it HST} observations and an example {\it SST} observation. 
In January 2005 observations were take with ACS (Program: 10452, PI: Beckwith) in 
the $F555W$ and $F814W$ filters. 
Star 1, the upper star of the two, is at RA 13:29:51.00 and Dec +47:11:26.95 
and Star 2 is at RA 13:29:51.02 and Dec +47:11:26.72.
Star 1 has V=$22.98\pm0.02$  and I=$20.93\pm0.01$ and 
the Star 2 has V=$23.92\pm0.03$ I=$23.18\pm0.02$ 
using DOLPHOT and correcting for galactic extinction.  
Observations were taken on 2008 March 31 with WFPC2 (Program: 11229, PI: Meixner) in the 
$F606W$ and $F814W$ filters.  
In these observations, we find Star 1 has V=$22.25\pm0.01$ mag and I= $20.44\pm0.01$ mag while 
Star 2 has V= $23.06\pm0.02$ mag and I=$22.98\pm0.06$.
Both stars appear to have brightened between 2005 and 2008.  

The most recent observation, a $F110W$ WFC3 IR image, was obtained on 2012 September 
04 (Program: 12490, PI: Koda).  
This falls on the epoch between when the candidate 
had faded below our threshold in the $R_c$ band and the end of our candidate selection period.  
We find a Vega {\it F110W} magnitude using aperture photometry of $20.79\pm0.01$ mag for the 
top source and $22.00\pm0.02$ mag for the bottom source.  
Both sources are also clearly visible in {\it F673N} and {\it F689M} images 
(Program:12762, PI: Kuntz) taken on 2012 April 10.  It is difficult to directly compare to 
the previous observations because of the filter differences.
However, we would expect one of these stars to no longer be detected if it was a failed SN.  
Therefore this observation supports eliminating this candidate.  
Since the last observation was obtained soon after the star met our selection criteria 
and well before the end of our survey period, new observations would be helpful to confirm 
its elimination. 

There are 16 epochs of IRAC data taken over 
10 years (Program: 159, PI:Kennicutt; Program: 30494, 
PI: Sugerman; Program: 40010, PI: Meixner; Program: 70207, PI: Helou; Program: 10136 and 
90240, PI: Kasliwal): 2 epochs in 2004, one in 2006 and 2007, 2 in 2008, 
10 in 2011, 6 in 2012 and 2 in 2013 and 2014. There is no source visible in any observation 
or any detectable variability.  
We calculate $3\sigma$ upper limits on the source flux of $0.182\pm.003$ mJy at 
$3.6\micron$ and $0.128\pm.003$ mJy at $4.5\micron$, which are not constraining in terms of the 
mid-IR luminosity.

\subsection{Former Candidate 4} 
The final former candidate was found in NGC~672 at RA 01:47:48.90 and Dec +27:25:28.82.  
This source was selected as a fading candidate in the $V$ and $R_c$ bands. Candidate 4 is 
not detected in the $U$ or $B$ bands and 
is blended with a brighter source in the $V$ and $R_c$ bands. 
Figure \ref{fig:672can_lc} shows the $V$ and $R_c$ band light curves for this candidate.  
We measure an $R_c$ band luminosity decrease of $1.3 \times 10^4 L_{\odot}$ and an initial 
magnitude of $R_c$ $\simeq$ 23.06 mag.  The $V$ band decreases in luminosity by 
$2.2 \times 10^4 L_{\odot}$, giving an initial magnitude of $V$ $\simeq$ 22.89 mag.
This source stayed undetected and within our criteria for candidacy for 
almost 2 years before LBT observations on 20 and 27 November 2014 show the source brightening, 
giving us solid grounds to reject it.  Figure \ref{fig:672can_image} shows selected observations.   
This source also shows the difficulties and ambiguities that can arise in areas 
where the image subtraction is less clean.

There are past observations of this source available from both {\it HST} and {\it SST}.  
The {\it HST} observations are in the $F336W$ filter from 2011 September 7 (Program: 12229, PI: Smith)
 and in the $F606W$ and $F814W$ filters from 2012 July 18 (Program: 12546, PI: Tully).  
After matching 
the astrometry of the HST observations to our LBT astrometry, we find no 
source in the $F336W$ data.  There is, however, a corresponding source in the $F606W$ and 
$F814W$ filters.  Using DOLPHOT to perform photometry and convert to 
Johnson filters and correcting for galactic extinction, we find Candidate 4 
has $23.39\pm0.01$ mag in the V band and $21.34\pm0.01$ mag 
in the I band. Figure~\ref{fig:672_arch} shows the  $F336W$ and $F606W$ observations 
and an example 3.6$\micron$ {\it SST} observation. 
There are archival IRAC SST observations 
for Candidate 4 taken in 2007 (Program: 40204, PI: Kennicutt) and 2014 
(Program: 10136, PI: Kasliwal).  This source is part of a star cluster and shows
 no variability in the SST data.  The 2007 and 2014 luminosities agree within 
the errors at $0.145\pm0.007$ mJy at $3.6\micron$ and $0.106\pm0.005$ mJy at $4.5\micron$.

\section{The Remaining Candidate}
Our survey period with 3 ccSNe resulted in one final failed SN candidate.  
This candidate shows a decrease in luminosity between the first and last 
observation of $\nu L_{\nu} \ge 10^4L_{\odot}$ (see Figure~\ref{fig:model2}).
Candidate 1 was observed to have a relatively 
stable luminosity for two epochs before it experienced an outburst and then faded. 
Archival {\it HST} observations show a clear source before the outburst. 
Archival {\it SST} data show that a long, slow mid-IR transient is associated with 
the source but the resolution makes it impossible to securely identify our source. 

The candidate was found in NGC~6946 at RA 20:35:27.56 and Dec +60:08:08.29.  
This candidate was first observed in May 2008 and then experienced an outburst in March 2009, after 
which it was no longer detected.  
The source was near a chip boundary and so was sometimes missed due to pointing variations, 
particularly in the first few years.  As a result, there are some epochs where 
we have $V$ band data but not necessarily $R_c$ band or $B$ band.
It was selected as a candidate in the $V$ and $R_c$ bands due to a large luminosity 
decrease.  With better sampling, this would have met our Lovegrove-Woosley model based criteria.  
The source's outburst caused one epoch in the V band to become masked.  We reprocessed this 
observation so that the source was no longer masked.   
In the discussion below, all luminosities and magnitudes are corrected for the 
significant Galactic extinction fo $E(B-V)$ = 0.30.
We measured a luminosity decrease in the $R_c$ band of $5.2 \times 10^4 L_{\odot}$ 
and an apparent magnitude of $R_c$ $\simeq $ 21.19 mag.  The $V$ band luminosity decrease
is $3.8 \times 10^4 L_{\odot}$ giving an initial $V$ band apparent magnitude 
of $V$ $\simeq $ 21.87 mag.  We estimate a $V-R$ color of 
$\simeq 0.68$ mag. 
Direct measurements of the initial magnitude of the candidate using DAOPHOT matches
these values to $\sim 0.1$ mag.

There is no source detected in the initial $U$ band observations and we find an apparent
 magnitude upper limit of $22.28\pm0.4$ mag. 
A $U$ band source with $20.0\pm0.1$ mag is visible in the $1\farcs9$ seeing 
observation that corresponds to the peak brightness in $V$ and $R_c$ bands.
The candidate was detected in November 2008 
and March 2009 in the $B$ band. The difference between the first and 
last $B$ band observations shows the source had an initial magnitude of $23.55\pm0.03$ mag. 
While the source increases to $22.30\pm0.04$ on 25 November 2008, it is unfortunately off the chip 
during the observation where the $V$ and $R_c$ bands peaked. 

\begin{figure}
\centerline{\includegraphics[width=3.5in]{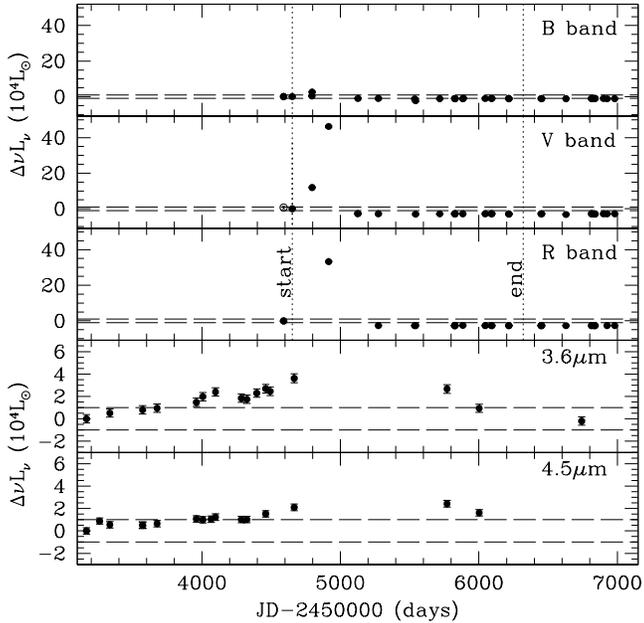}}
\caption{The $B$, $V$ and $R_c$ band differential light curves for Candidate 1 in 
NGC~6946.  The open circle in the $V$ band light curve was an observation that fell just 
outside our quality criteria that was later added as a check on the measurements. 
The vertical axis is in units of $10^4L_{\odot} (\nu L_{\nu})$ and has 
been normalized to the first observation so that the luminosity difference between 
the first and last observations can be easily seen.  A change in luminosity by 
$10^4L_{\odot}$ in either direction, as indicated by the horizontal lines, 
would lead to the source being selected as a candidate.
The bottom two panels are the $3.6\micron$ and $4.5\micron$ {\it SST} archival light curves 
normalized to the first epoch and on a different y-axis scale.
}
\label{fig:6946can_lc}
\end{figure}

\begin{figure*}
\centerline{\includegraphics[width=7.0in]{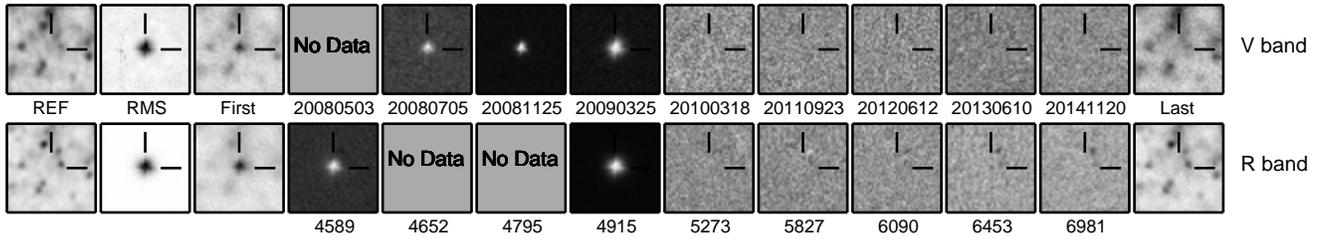}}
\caption{Select $V$ and $R_c$ band observations for Candidate 1 in 
NGC~6946.  We have 19 epochs for this galaxy and do not show them all.  The 
selected observations give a clear picture of the source's variability. 
The ``First'' observation in the $V$ band ($R_c$ band) is on 5 July 2008 (3 May 2008) and 
the ``Last'' observation is on 20 November 2014.  
The format is the same as in Figure~\protect\ref{fig:ex_cut}.
}
\label{fig:6946can_image}
\end{figure*}

\begin{figure}
\centerline{\includegraphics[width=3.5in]{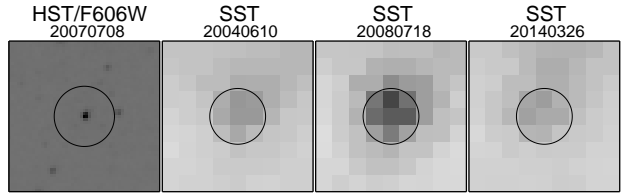}}
\caption{Archival observations of Candidate 1.  The $F606W$ {\it HST} observation and 
the first, brightest and last 3.6$\micron$ {\it SST} observations.  The archival 
images are labeled with the date of the observation. The circle marks the Candidate 
location and has a 1 arcsecond radius.  The images are 5 arcseconds on a side. 
}
\label{fig:6946_arch}
\end{figure}

Figure \ref{fig:6946can_lc} shows the $B$, $V$ and $R_c$ light curves along with archival 
{\it Spitzer Space Telescope} $3.6 \micron$ and $4.5 \micron$ light curves that will be discussed later.  
We can see a source 
that is detected in two epochs at the beginning of the survey.  
In the $R_c$ band, we observed the source on two successive nights, 
3/4 May 2008 (the light curve points overlap in Figure~\ref{fig:6946can_lc}). 
NGC~6946 was also observed in the $V$ band on 4 May 2008 with data quality 
just outside our analysis criteria.  
We include this observation as an open point in the light curve and used this 
for our measurement of initial luminosity so that it could be compared to the R band on 
that same date.  We measured the 
differential flux with simple aperture photometry as a comparison to the ISIS estimates 
and found good agreement. 

The peak brightness we observe is on 25 March 2009 for both the $V$ and $R_c$ bands.   
We measure $V$ $\simeq $ 18.17 mag ($\nu L_{\nu} = 1.15 \times 10^6 L_{\odot}$) and $R_c$ $\simeq $ 17.58 mag
($\nu L_{\nu} = 14.34 \times 10^6 L_{\odot}$).  
After this peak,
the source was not detected in any band for the remainder 
of our survey, with the last observation for this galaxy on 20 November 2014.
Figure \ref{fig:6946can_image} shows select observations for both the $V$ and $R_c$ bands. 
 The candidate is clearly detected 
in the first epoch, experiences an outburst and is not visible on or after 20 October 2009. 

We found no other references to this outburst.  There is a cataloged GALEX UV source close to 
its position, however there is also a 21.71 mag $U$ band source within 4 arcseconds 
from our candidate that is likely the GALEX source. 
The detection of the candidate two nights in a row at a relatively unchanged luminosity 
in May 2008 shows 
that the source was present and relatively stable at the start of the survey.  If this 
outburst was a nova or some other type of stellar variability, we expect that the star would 
not have been seen earlier, never fully disappeared, or should have returned.   
Based on the LBT data, this is a promising candidate. 

There are archival observations of this source from both {\it HST} and the 
{\it SST} and Candidate 1 is easily identified in the observations 
from both telescopes.  There is a single 
epoch of {\it HST} data from 8 July 2007 in the $F606W$ and $F814W$ WFPC2 filters 
(Program: 11229, PI:Meixner).  
Using the DOLPHOT software package \citep{dolphin2002} to perform photometry and convert to 
Vega magnitudes and correcting for Galactic extinction, we find Candidate 1 
magnitudes of $22.15\pm0.01$ in the V band and $20.28\pm0.01$ mag in the I band. 
There are 16 IRAC observations at $3.6\micron$ and at $4.5\micron$ 
(Program:159, PI: Kennicutt; Program: 
3248, 20256 and 30292, PI: Meikle; Program: 30494, PI: Sugerman; Program: 40619, PI: Kotak; 
Program: 40010, PI: Meixner; Program: 80015, PI: Kochanek; Program: 10136, PI: Kasliwal).  
We preformed image subtraction on these observations using the same procedures as for our LBT 
data and produced the 
light curves shown in Figure~\ref{fig:6946can_lc} with our LBT light curves. 

 Unlike the quick optical brightness decrease seen after the outburst, the increase and decrease of brightness 
in the mid-IR is very slow and continuous through the last observation at $3.6\micron$ on 2014 March 26.  
Unfortunately, there is a gap in the mid-IR data when we see the outburst in the optical and the date and shape 
of the mid-IR peak are unknown. 
 Figure~\ref{fig:6946_arch} 
shows the $F606W$ {\it HST} image and 3 of the 3.6$\micron$ {\it SST} observations (first, 
brightest and last).  The archival images are labeled with the date of the observation.  
While there is some luminosity visible in the area of the source in the {\it SST} observations after it 
has vanished in the LBT data, and Candidate 1 could be a dusty stellar transient,
it is part of some extended emission and the resolution is insufficient to securely detect the 
individual source seen in the LBT and {\it HST} observations.  It could also still be fading in the 
$SST$ bands.  Additional observations, by both {\it SST} and {\it HST} are needed 
to determine the nature of this candidate since its outburst.  
We conclude that Candidate 1 is a viable candidate for a failed SN.  

We fit the HST, LBT (including the U band upper limit) and SST
data from near 8 July 2007, where we measure [3.6]=$17.51\pm0.05$ mag 
and [4.5]=$17.24\pm0.05$, using Solar metallicity \citet{castelli2004} 
model atmospheres obscured by
circumstellar silicate dust, and using DUSTY (\citealt{Ivezic1999}; \citealt{elitzur2001}) to model the
radiation transport.  We assumed 10\% photometric errors to
compensate for mixing data from  modestly different dates.  If we include
no circumstellar dust we find $L_* \simeq 10^{5.25\pm0.02}L_\odot$
and $T_* \simeq 3600 \pm 200$~K (nominally at 90\% confidence)
but with a best fit of $\chi^2=40.5$ for 7 degrees of freedom.
If we include the dust, we find $L_* \simeq 10^{5.11\pm0.08} L_\odot$
and $T_* \simeq 4700 \pm 800$~K with $\log \tau_V \simeq 0.4_{-0.2}^{+0.1}$
of dust.  The models with dust fit the data well, with $\chi^2=8.1$ for
6 degrees of freedom.  If we map either of these models onto the
end points of the Solar metallicity PARSEC isochrones \citep{bressan2012},
they correspond to roughly $18M_\odot \la M_* \la 25M_\odot$ stars with the
cooler, no dust models better matching the very end points of the
isochrones.  Thus, Candidate 1 appears to have properties that
put it almost exactly in the mass range corresponding to the
red supergiant problem.

\section{Limits on the Rate of Failed SNe}
If the rate of core collapses in the sample is $r$ and the failed fraction 
is $f$, then the expected number of ccSNe and failed SNe are 
$N_{SN} =r(1-f)t_{SN}$ and $ N_{FSN} =rft_{FSN}$, 
where $t_{SN}$ is the period over which we search for SNe and $t_{FSN}$ is the 
failed SNe candidate selection period.  
The two survey times $t_{SN}$ and $t_{FSN}$ need not be the same because the SN 
surveys of these galaxies are not solely dependent on the LBT data. 
To determine the constraints on the failed fraction, we assume ideal 
detection efficiency and 
use Poisson probability distributions and marginalize over the 
core-collapse rate $r$.
Taking a Bayesian approach and using a logarithmic prior of $P(r) \propto (1/r)$ for 
the rate and a uniform prior $P(f) \propto constant$ for the failed fraction, we find the probability 
of getting a fraction, $f$, of failed SNe is proportional to, 
\begin{multline}
 P(f) \propto  (1 - f)^{N_{SN}}      \cdot \\
f^{N_{FSN}} (f (t_{FSN} - t_{SN}) + t_{SN})^{(-N_{SN} - N_{FSN})},
\end{multline}
where $N_{SN}$ and $N_{FSN}$ are the observed number of ccSNe and failed SNe, respectively.  If we 
use a uniform prior $P(r)\propto constant$, we find the probability is proportional to
\begin{multline}
P(f) \propto  (1 - f)^{N_{SN}} \cdot \\ 
f^{N_{FSN}} (f (t_{FSN} - t_{SN}) + t_{SN})^{(-1-N_{SN} - N_{FSN})}.
\end{multline}
We have dropped constant terms and the normalizations of the expressions can be found 
by requiring $ \int^1_0 P(f) df \equiv 1$.  There are explicit expressions in terms of 
hypergeometric functions, but they are not illuminating.  The difference in the 
priors appears as a change in the exponent of the third term.  If we set the survey 
times to be identical, the resulting binomial distribution, $P(f) \propto (1-f)^{N_{SN}} f^{N_{FSN}}$, 
does not depend on the prior on $r$.  To constrain the fraction of
failed SNe, we must choose a SN survey period, $t_{SN}$, with $N_{SN}$ known ccSNe, 
and a failed SN survey period, $t_{FSN}$, with $N_{FSN}$ failed SN candidates.  
To summarize, the number of ccSNe constrains $r(1-f)$, while the number of 
failed SNe constrains $rf$ and by marginalizing over $r$, we obtain the constraint
on $f$.

\begin{figure}
\centerline{\includegraphics[width=3.5in]{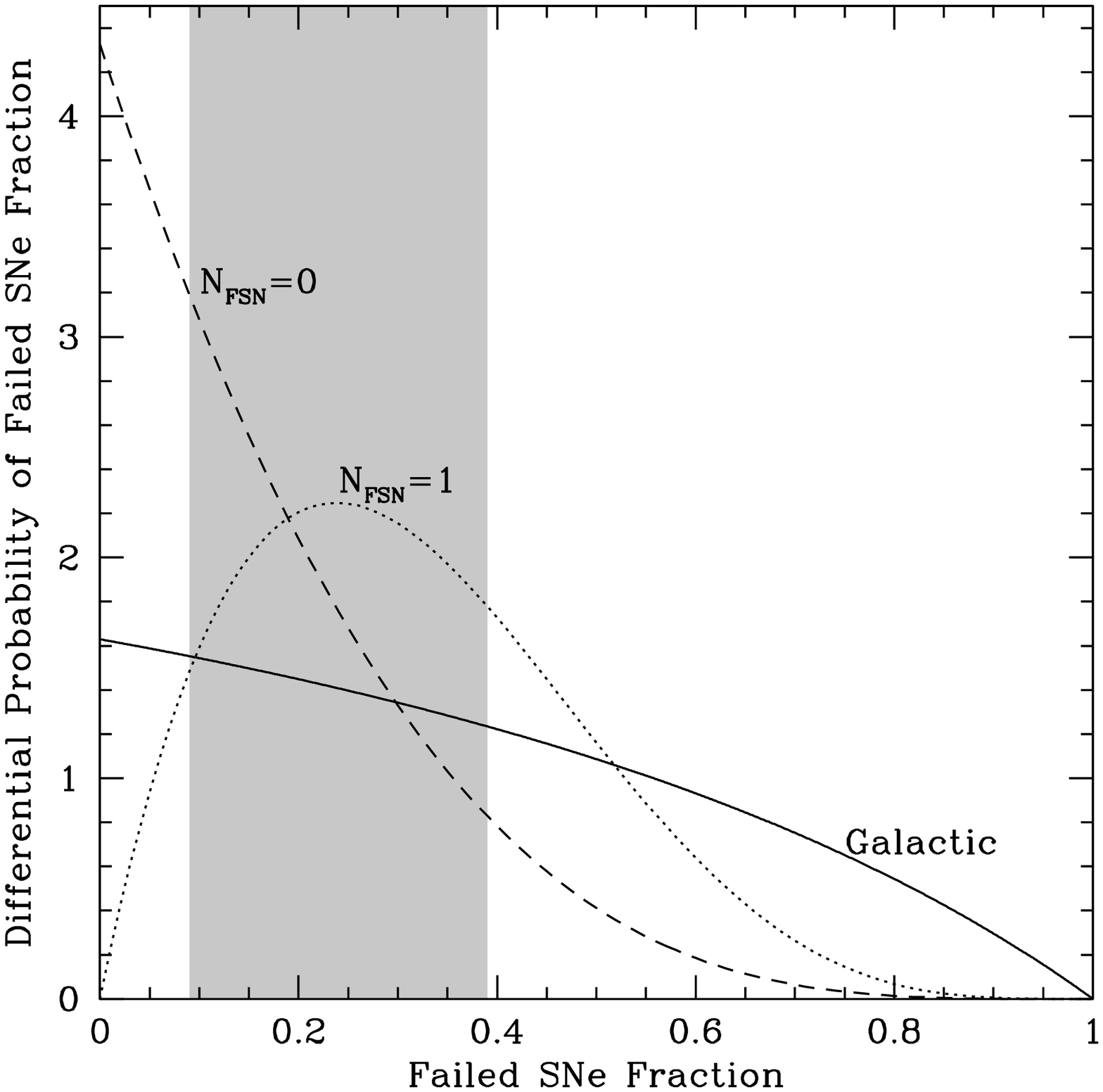}}
\caption{Probability dP$/$d$f$ for the fraction of ccSNe that are failed SNe if 
$N_{FSN}=0$ (dashed) or 1 (dotted) failed SN are found in our sample. 
The solid line shows the constraints from combining the historical Galactic SN rate 
with the non-detection of neutrinos from a Galactic ccSN \protect\citep{adams2013}. 
\protect\citet{horiuchi2011} suggests $f \la 0.5$ given the comparison between the SN rate and the star 
formation rate.  The red supergiant problem is solved if $f \sim 0.2 - 0.3$ 
 and the black hole mass function suggests $0.092 < f < 0.39$ \protect\citep{kochanek2014b} 
as shown by the shaded region. 
}
\label{fig:limits}
\end{figure}

\begin{table*}
\caption{Fraction of Failed SN} 
\label{tab:frac} 
\begin{tabular}{ccccccccc}
\hline \hline
   $t_{SN}$ & $\#$ SN & $\#$ Failed SN & \multicolumn{2}{c}{Upper Limit} &
 \multicolumn{2}{c}{Range} &   \multicolumn{2}{c}{Median} \\
 (years) & $N_{SN}$ & $N_{FSN}$ & log &  const & log &  const
 & log &  const \\
\hline
14.0 & 12 & 0 & 0.345 & 0.373 & & & & \\
24.0 & 18 & 0 & 0.462 & 0.522 & & & & \\
4.0 & 3 &  0 & 0.438 & 0.438 & & & & \\
5.0 & 4 &  0 & 0.404 & 0.413 &  &  & & \\
5.0 & 4 & 1 & 0.550 & 0.558 & 0.073 - 0.620 & 0.075 - 0.627 & 0.296 & 0.303\\
\hline
\multicolumn{9}{p{0.7\textwidth}}{The upper limits and ranges are at a 90\% confidence level.  The columns labeled 
``log'' have a logarithmic prior of $(1/r)$ on the unknown core-collapse rate while those labeled
``const'' have a uniform prior. }
\end{tabular}
\end{table*}

We could estimate $\langle N_{SN} \rangle$ from the historical record in the Sternberg
Astronomical Institute supernova catalog.  The sample galaxies had 26 probable ccSNe since 1900
along with 4 Type~Ia SNe \citep{tsvetkov2004}.  If we calculate the typical rate using the time period
from 1 January 2000 to 1 January 2014 there are 12 ccSNe ($t_{SN}=14.0, N_{SN}=12$), implying a rate of
$r_{ccSN} = 1.01$ year$^{-1}$ with a $90\% $ confidence range of $0.55 < r_{ccSN} < 1.39$.
If we go back to 1 January 1970, there were 18 ccSNe ($t_{SN}=44.0, N_{SN}=18$), implying a rate of
$r_{ccSN}=0.41$ and a 90\% confidence range of $0.28 < r_{ccSN} < 0.61$,
although it is also clear that the samples are almost certainly becoming
incomplete on these longer baselines.
However, we know we did not miss any new SNe during our survey.  
The survey data that is analyzed in this work is from the beginning of the survey in 
2008 until the end of 2013. 
As discussed previously, there were 3 ccSNe during our candidate selection period ($t_{SN}=4.0, N_{SN}=3$).  
There was an additional ccSN between the candidate selection period and the end of 2013, 
the type II-P SN~2013ej in NGC~628 in July 2013 \citep{valenti2013}.  This gives a total 
of 4 ccSNe ($t_{SN}=5.0, N_{SN}=4$) and a rate during the overall survey of $r _{ccSN}= 0.80$ year$^{-1}$, 
which is consistent with the recent historical record.  We will use this case as our fiducial example. 

For the failed SNe, we use the time from the second epoch to the last 
observation of our candidate selection period to find a galaxy-averaged baseline of 
$ t_{FSN}=4.0$ yrs.  We use the second observation 
to begin our baseline because we require the star to be observed for two epochs before it 
fades to eliminate the possibility of a false-positive caused by a 
nova in the first epoch.  If we include the first epoch, the average baseline is $4.6 \pm0.3$ years.
Here we simply use the average of the survey periods for each galaxy.  While there 
is some spread (1.7--4.8  years), how we do the average is a negligible source of uncertainty.  
The average masked fraction for the galaxies is $m=(3.9\pm1.1)$\% which we can view as a 
reduction in our effective survey time, $t_{FSN} \rightarrow (1-m)t_{FSN}$.  
However, we monitor saturated stars that are masked, 
effectively reducing the masked fraction, and the uncertainties introduced by the small masked 
fraction are negligible compared to our present statistical uncertainties and will be ignored. 
The primary systematic uncertainty is the fraction of failed SNe we would 
miss because the star is heavily obscured by interstellar dust.  This 
will be empirically explored by our ability to recover heavily extincted 
SN progenitors like that of SN~2009hd.  

In \S7 we were left with one possible candidate failed SN.  Further 
monitoring and new {\it HST} observations should clarify its status, but for now we 
consider the cases of $N_{FSN}$= 0 and 1.  Table \ref{tab:frac} displays the resulting constraints on 
the fraction $f$ of massive stars that experience a failed SN.  We calculate the limits using both 
a logarithmic prior of $P(r) \propto (1/r)$ (columns labeled ``log'') and a uniform prior 
$P(f) \propto constant$ (columns labeled ``const'') for the underlying core-collapse rate $r$.  
For simplicity, we first consider the $N_{FSN}=0$ case, which occurs if the remaining candidate 
is ruled out.  For our fiducial case ($t_{SN}=5.0, N_{SN}=4$) this leads to 
a (one-sided) 90\% confidence upper limit of $f \le 0.40 (0.41)$ for the logarithmic 
(uniform) prior on $r$.  Changing the assumed supernova survey period shifts the limits by 
roughly 10\%, as shown in Table~\ref{tab:frac}.  
With $N_{FSN}=1$ the (one-sided) 90\% confidence upper limit rises to $f \le 0.55$ (0.50).  
Of course, with a detection we would really have a measurement, 
and for $N_{FSN}=1$, the median fraction is $f \simeq 0.30$, with symmetric 90\% confidence limits of 
$0.07 \le f \le 0.62$. Figure~\ref{fig:limits} shows the differential probability for the failed SNe fraction $f$ 
for both of these cases.

\section{Discussion}
This initial survey suggests several improvements for our next analysis.  
First, we plan to revise the masking procedure.  
The physical masking of the images before subtraction for reasons other 
than CCD defects needs to be reduced or removed entirely, with masking of saturated regions 
taking place after image subtraction to reduce the number of false positive 
detections near the edges of masked regions.  
One option to improve our candidate selection is to require the candidate to be 
securely detected for a longer period.  There would be no constraints on 
the variability of the source since the behaviors of core-collapse progenitors are not 
well understood.  
This change would ensure we would be able to characterize any candidate reasonably 
well and further ensure that we were not catching the end of a nova or outburst event. 
The main improvement to future limits from this survey will come from an 
increased time baseline and SN sample.  We already have an additional two years that can be 
added to our survey selection period and an additional ccSN 
(SN~2014bc in NGC~4258; \citealt{smartt2014}). 

Aside from our approach, the only other prospect of detecting a failed SN is to 
observe neutrinos or gravitational waves from a core collapse in our Galaxy.  This would 
have the advantage of directly probing the collapse to a black hole, but the disadvantage 
of a very low rate.  
If the Galactic SN rate is one every 50-100 years, and the failed SN fraction 
is $f = 0.3 $, there is a failed SN in the Galaxy only once every 150 to 300 years. 
\citet{adams2013} combined the observed SN rate of the galaxy with 
the absence of any neutrino detections (\citealt{alex2002}; \citealt{ikeda2007}) of a failed 
SN over the last $\sim$3 decades to estimate that $f \le 0.69$ at
90\% confidence.  Figure~\ref{fig:limits} shows how these constraints compare to 
those in this work.  Our much larger sample of massive stars has much more 
power in constraining this rare phenomenon. 

Other attempts to estimate the failed SNe fraction come from cosmic measurements.  
The non-detection of the diffuse SN background gives an upper limit of 
$f \le 0.50 - 0.75$ \citep{lien2010}.
Comparing massive star formation rates to SNe rates along with the lack of a detection 
of a diffuse neutrino background from SNe, \citet{hopkins2006} estimates that the fractional 
failed SN rate cannot much exceed $f=0.5$.
Horiuchi et al. (2011, also see 2014), finds a discrepancy between the cosmic SNR and the massive 
star formation rate which allows for a fractional failed SN rate up to $\sim$0.5 
(but see \citealt{botticella2012}).  Our limits are consistent with these estimates.  
The black hole mass function can be explained by a failed SN rate of $0.09 \le f \le 0.39$ 
 \citep{kochanek2014b} and such failure rates are consistent with studies of the ``explodability'' of massive 
stars (\citealt{oconnor2011}; \citealt{ugliano2012}; \citealt{ondrej2014}). 

While the primary goal of this survey is to better understand the fates of massive 
stars, it enables a broad range of other scientific explorations.  
For example, we should be able to obtain $UBVR$ photometry of all future ccSN 
progenitors in these galaxies.  Our sensitivity in
searching for the progenitor of a particular SN, where we know exactly where
to conduct the search, will be considerably better than in the search for failed SNe.  
We estimate that we should
be able to measure the properties of almost all SN progenitors because, in this
case, we can compare all the stacked data prior to the explosion to as many
epochs as needed once the supernova has faded rather than needing to follow
the evolution of the progenitor luminosity.  This can be seen in the high 
signal-to-noise detection of the progenitor of SN~2011dh 
(see Figures~\ref{fig:2011dh_lc} and \ref{fig:2011dh_image}). 
The data also open a new field, progenitor 
variability, whether due to stellar activity or binarity (ellipsoidal variations or 
eclipses).  This was demonstrated in \citet{dorota2012a}, where we used LBT to determine 
the variability of the progenitor of SN~2011dh.  
The data are also being used to identify LBVs \citep{grammer2015} 
 and to help characterize dusty evolved stars similar to $\eta$ Carina or 
 IRC+10420 \citep{khan2014}. 

Our cadence and long baseline makes this survey ideal of studying long-term 
variables such as Cepheids.  We have completed
Cepheid studies for M81 and NGC~4258, determining the distances to these 
galaxies and exploring the dependence of Cepheid brightness on composition 
(\citealt{gerke2011}; \citealt{fausnaugh2014}).  Our large field of view and long 
baseline allows us to identify Cepheids throughout the galaxy over a range of metallicities
 and periods $P \ga 10$ days.  
 Other possibilities include the study of the long term variability of a large 
population of red supergiants and searches for massive eclipsing binaries.  For example, 
\citet{prieto2008a} presents the discovery of an eclipsing binary in the Dwarf Galaxy Holmberg IX.  
Finally, this survey can also be utilized to search for other rare phenomena 
that produce an optical signature, such as stellar mergers (see \citealt{kochanek2014m}).  
 
We should note the LBT/LBC is the best telescope/instrument for this survey.  
The wider field of view of Hyper Suprime-Cam on Subaru is only important for M31 and M33, but the 
binocular mode matters more for all other galaxies.  
HST could carry out the survey to 
 a distance of 30 Mpc, but the costs would exceed even the scope of an HST Legacy Program. 
LSST could carry out a similar study in the South, but it would require a dedicated sub-survey 
(the standard exposures are far too shallow) and 
 would have a galaxy sample with a lower ccSN rate. 
 Moreover, if LSST starts science observations in 2022 as 
scheduled, it would have a data set comparable to that from our LBT survey, which is 
currently planned to continue till 2017, in about 2035.  
The most promising possible extension to this survey would be with 
WFIRST, although the lack of bluer optical bands will limit the characterization of the stars.

\section{Conclusions}
We have analyzed the first 4 years ($t_{FSN}=4.0$) of the LBT search for failed SNe.  
We observed 27 nearby galaxies in the $U_{spec}$, $B$, $V$, and $R$ bands 
multiple times a year to monitor $\sim10^6$ red supergiants.  We analyzed many target 
sources in our survey and are left with 1 final failed SN candidate.  
Candidate 1 in NGC~6946 had an initial apparent magnitude of $R_c$ $\simeq $ 21.89 mag 
($\nu L_{\nu} = 2.7 \times 10^4 L_{\odot}$).  It was observed to have a relatively 
stable luminosity for two epochs before it experienced an outburst and then faded. 
Archival {\it SST} data show a long, slow mid-IR transient is associated with 
the source but the resolution makes it impossible to securely identify our source.  
Follow up observations with {\it HST}/{\it SST} and continued monitoring should help better 
determine the nature of this candidate. Fitting the available data with stellar models and 
mapping those results onto isochrones imply a red star with a mass of $18M_\odot \la M_* \la 25M_\odot$, 
placing this candidate in the mass range of the red supergiant problem.

We use the known ccSNe in these galaxies as tests of our approach.  
Of the 3 massive stars known to have died during our survey period, SN~2009dh, SN~2011dh and 
SN~2013fh, we select SN~2011dh as a star which has died.  The other 2 SNe have not faded 
sufficiently to be found as vanishing stars. 
In the future, we will experiment with adding fake candidates.  We have not done so 
as yet, instead focusing on whether we find the deaths of stars in successful SNe, because it 
is not entirely clear what fake signal to inject.  There is a need for additional studies 
on the observational signatures of failed SNe such as that by \citet{lovegrove2013} and 
\cite{piro2013} to allow such calibration studies.

Because we have a remaining candidate failed SN, we consider the constraints on the fraction of 
core-collapses that lead to a failed SNe if our sample contains 0 or 1 failed SN.  
Using a log prior for our fiducial case for the SN survey ($N_{SN}=4$ discovered over 
$t_{SN}=5.0$ years), 
we find an upper limit on the fraction of massive stars that experience a failed SN 
of $f \simeq 0.40$ at 90\% confidence if we ultimately reject our remaining candidate.  
If we have discovered a failed SN, the median fraction is $f \simeq 0.30$ 
with symmetric 90\% confidence limits of $0.07 \le f \le 0.62$.

The survey is continuing, and we are planning on a minimum survey duration 
of 10 years.  If we conservatively constrain the ccSN rate using the observed numbers 
from 2000 to 2014 ($t_{SN}=14,N_{SN}=12$), then if we find no failed ccSNe, the 90\% (95\%)
confidence upper bounds on $f$ is $f < 0.18$ ($f < 0.31$).  Similarly, the probabilities 
of finding a failed SN are 63\%, 88\% and 97\% for $f=0.1, 0.2$ and $0.3$, respectively.  
These estimates are somewhat conservative because we did not take into account 
the improvements in the estimate of the ccSNe rate given the extended temporal baseline.  
This is a challenging experiment, but it will set interesting limits and the 
prospects of a revolutionary discovery are good.  Of course, if Candidate 1 survives,
the revolutionary discovery has already occurred.

\section*{acknowledgments}
We thank T.A.~Thompson and J.F.~Beacom for comments and discussions.
We thank S. Adams for helpful discussions and providing the ``Galactic" curve 
in Figure~\ref{fig:limits}. We thank the OSURC queue observers and the support
staff at LBT for helping us obtain these observations. 
We also thank Stephen Smartt, the referee, for many helpful suggestions 
which strengthened the paper.
Observations have been carried out using the Large Binocular Telescope at
Mt. Graham, AZ. The LBT is an international collaboration among institutions
in the United States, Italy, and Germany. LBT Corporation partners are the
University of Arizona on behalf of the Arizona university system; Istituto
Nazionale di Astrofisica, Italy; LBT Beteiligungsgesellschaft, Germany,
representing the Max-Planck Society, the Astrophysical Institute Potsdam, and
Heidelberg University; the Ohio State University; and The Research
Corporation, on behalf of the University of Notre Dame, University of
Minnesota, and University of Virginia.
Based on observations made with the NASA/ESA Hubble Space Telescope, obtained 
from the data archive at the Space Telescope Science Institute. STScI is 
operated by the Association of Universities for Research in Astronomy, Inc. 
under NASA contract NAS 5-26555.
This work is based in part on observations made with the Spitzer Space Telescope, 
which is operated by the Jet Propulsion Laboratory, California Institute of 
Technology under a contract with NASA.


\end{document}